\documentclass[10pt,journal,compsoc]{IEEEtran}
\ifCLASSOPTIONcompsoc
  \usepackage[nocompress]{cite}
\else
  \usepackage{cite}
\fi
\ifCLASSINFOpdf
\else
\fi

\usepackage{wrapfig}
\usepackage{adjustbox}
\usepackage{amsmath,amssymb,amsfonts}
\usepackage{algorithmic}
\usepackage{graphicx}
\usepackage{textcomp}
\usepackage{amsmath,amssymb,amsfonts}
\usepackage{algorithmic}
\usepackage{graphicx}
\usepackage{textcomp}
\usepackage{subcaption}
\usepackage{epsfig}
 
\usepackage{array}
\usepackage{multirow}
\usepackage{adjustbox}
\usepackage{booktabs,siunitx}
\usepackage{stfloats}
\usepackage{algorithm}
\usepackage{url}
\usepackage{enumitem}
\usepackage{interval}
\usepackage{tabularx}
\usepackage{threeparttable}
\usepackage{verbatimbox}
\usepackage{makecell}
\usepackage[utf8]{inputenc}
\usepackage[T1]{fontenc}
\usepackage{lineno}

\begin{document}
%
\title{Transformer-based Self-supervised Multimodal Representation Learning for Wearable Emotion Recognition}
%
%
%
%
\author{Yujin~Wu,
        Mohamed~Daoudi,~\IEEEmembership{IEEE Senior},
        Ali~Amad
        
\IEEEcompsocitemizethanks{\IEEEcompsocthanksitem Y. Wu is with Univ. Lille, CNRS, Centrale Lille, UMR 9189 CRIStAL, Lille, F-59000, France.
\protect\\
E-mail: yujin.wu.etu@univ-lille.fr
\IEEEcompsocthanksitem M. Daoudi is with Univ. Lille, CNRS, Centrale Lille, Institut Mines-Télécom, UMR 9189 CRIStAL, F-59000 Lille, France, and IMT Nord Europe, Institut Mines-Télécom, Centre for Digital Systems.
\protect\\
E-mail: mohamed.daoudi@imt-nord-europe.fr
\IEEEcompsocthanksitem A. Amad is with the Lille Neuroscience $\&$ Cognition U1172 INSERM labotory, University of Lille, 59000 France.
\protect\\
E-mail: ali.amad@univ-lille.fr
}
}

%
%

\markboth{Journal of \LaTeX\ Class Files}%
{Shell \MakeLowercase{\textit{et al.}}: Bare Demo of IEEEtran.cls for Computer Society Journals}
%



\IEEEtitleabstractindextext{%
\begin{abstract}
Recently, wearable emotion recognition based on peripheral physiological signals has drawn massive attention due to its less invasive nature and its applicability in real-life scenarios. However, how to effectively fuse multimodal data remains a challenging problem. Moreover, traditional fully-supervised based approaches suffer from overfitting given limited labeled data. 
To address the above issues, we propose a novel self-supervised learning (SSL) framework for wearable emotion recognition, where efficient multimodal fusion is realized with temporal convolution-based modality-specific encoders and a transformer-based shared encoder, capturing both intra-modal and inter-modal correlations. 
Extensive unlabeled data is automatically assigned labels by five signal transforms, and the proposed SSL model is pre-trained with signal transformation recognition as a pretext task, allowing the extraction of generalized multimodal representations for emotion-related downstream tasks. 
For evaluation, the proposed SSL model was first pre-trained on a large-scale self-collected physiological dataset and the resulting encoder was subsequently frozen or fine-tuned on three public supervised emotion recognition datasets. 
Ultimately, our SSL-based method achieved state-of-the-art results in various emotion classification tasks. Meanwhile, the proposed model was proved to be more accurate and robust compared to fully-supervised methods on low data regimes.
\end{abstract}

}

\maketitle

\IEEEdisplaynontitleabstractindextext

\begin{figure*}[!htb]
\centering
\includegraphics[scale = 1.1]{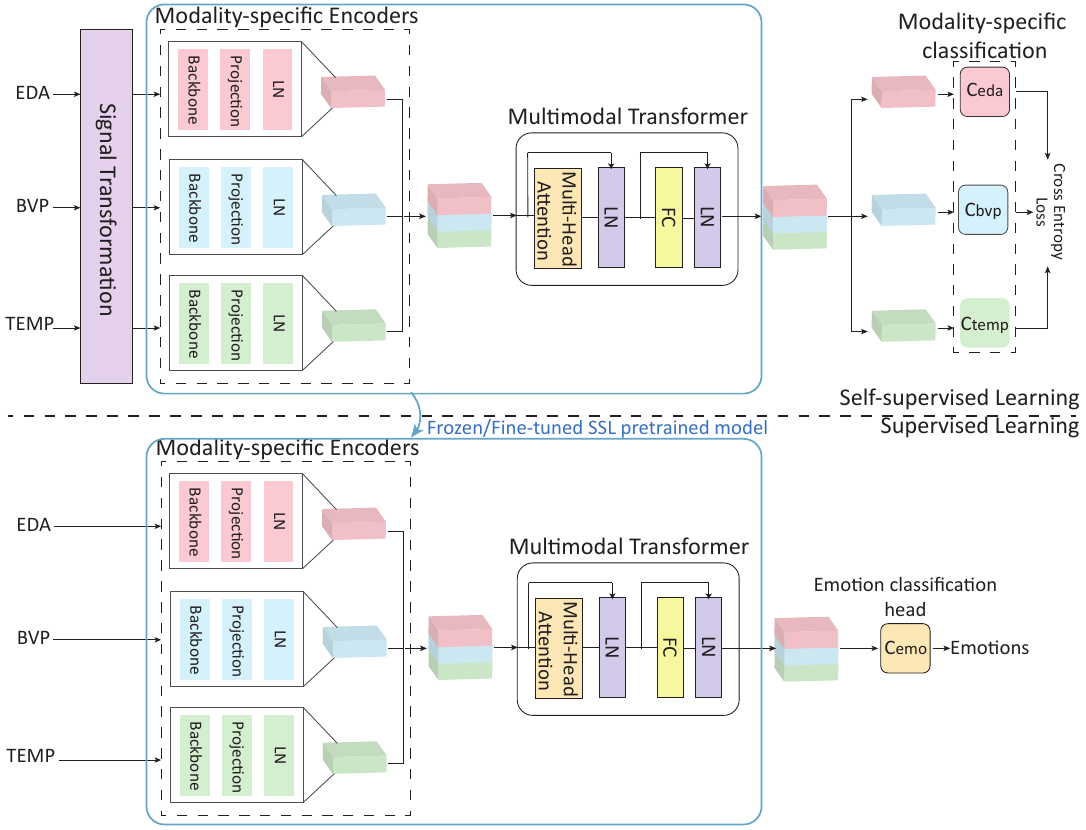}
\caption{Overview of our self-supervised multimodal representation learning framework. 
The proposed SSL model is first pre-trained with signal transform recognition as the pretext task to learn generalized multimodal representation.
The encoder part of the resulting pre-trained model is then served as a feature extractor for downstream tasks which is frozen or fine-tuned on the labeled samples to predict emotion classes.
}
\label{Fig1:Overview}
\end{figure*}

%
\IEEEpeerreviewmaketitle
\IEEEraisesectionheading{\section{Introduction}\label{sec:introduction}}
\IEEEPARstart{E}{motions} are sets of complex physiological, cognitive and behavioral responses that are triggered by internal or external stimuli. 
Emotion recognition is an emerging field of research which attempts to empower computers with the ability to infer human emotions. 
In recent years, it has been employed in several practical scenarios such as automated driver assistance \cite{driver_stress}, health care \cite{ayata_emotion_2020}, social communication \cite{DAI2015777}, etc, the majority of which are based on physical or physiological indicators of the human body. 
In contrast to physical signals such as facial expressions\cite{JAIN201969} and speech \cite{8682283}, physiological responses under certain emotional states are involuntary and therefore provide more objective decisions for identification systems \cite{WANG202219}. 
The physiological modalities primarily consist of Electroencephalography (EEG) signals and a series of peripheral signals. 
However, the acquisition of EEG data is challenging for implementation in real-life scenarios. With the advance of non-invasive technologies, emotion recognition methods based on multiple peripheral signals captured by smartphones/wearable watches have attracted some attention. 
Most recent researches focus on deep neural networks, which can automatically extract complex patterns from multimodal signals.
However, given that most of them are trained in a supervised manner, it is challenging to obtain generalizable models using limited labeled data, especially in daily life, where standard protocols for obtaining accurate emotion labels are not yet well defined. 
Besides, each specific supervised task requires training the deep model from scratch and its knowledge transfer ability on other tasks is not satisfactory \cite{ecg_ssl}. Self-supervised learning (SSL), as an emerging learning paradigm, eliminates the need for extensive manual labeling and has demonstrated comparable or even superior performance to supervised learning methods in areas of computer vision (CV), natural language processing (NLP). Several SSL-based efforts ~\cite{Banville_2021, jiang_eeg, cheng_eeg} have been done for emotion recognition using EEG signal, but they are not suitable for practical scenes. 
Only one work \cite{sigrep} targeted low-frequency wearable peripheral signals, but they ignored the correlation between multimodal signals. 
In this paper, we propose a self-supervised multimodal representation learning approach for wearable emotion recognition based on peripheral physiological signals. The first stage is model pre-training with the pretext objective of signal transformation classification, where a large amount of unlabeled multimodal data are automatically assigned labels through a series of transformations. 
Considering the heterogeneity of multimodal signals, temporal convolution-based modality-specific encoders are first employed separately on the transformed unimodal data to extract low-level features, followed by a transformer-based shared encoder deployed to aggregate unimodal features, enabling the modeling of complementary and collaborative properties between multimodal signals. 
Finally, modality-specific signal transformation recognition is performed to learn effective multimodal representations for downstream tasks that are robust to perturbations in magnitude or temporal domains. 
The second stage is supervised emotion recognition, where the SSL pre-trained encoder part is retained as a feature extractor to obtain generalized multimodal representations for classification. The overview of the proposed approach is illustrated in Fig. \ref{Fig1:Overview}.
To validate the effectiveness of our method and the knowledge transferability across different datasets,
we pre-trained the proposed model on a large-scale unsupervised emotion dataset PRESAGE collected in unrestricted real-life scenarios and evaluated its performance on three public emotion recognition datasets. Overall, our contributions can be summarized as follows:
\begin{itemize}
  \item We proposed a novel self-supervised learning (SSL) framework to learn generalized representations from a large number of unlabeled samples to cope with the overfitting problem on small-scale physiological data.
  \item We adopted an intermediate fusion strategy based on temporal convolution and transformer, capable of modeling both the heterogeneity and cross-modal correlation of physiological signals to effectively fuse multimodal data.
  \item We outperformed state-of-the-art supervised or self-supervised learning-based approaches in various emotion-related classification tasks involving mental stress, affective states, arousal, and valence. Moreover, our model was proven to be more accurate and stable on limited labeled data than fully-supervised models. In addition, multiple ablation studies have been performed to investigate the effectiveness of our method.
 \end{itemize}
\section{Related Work}
\label{sec:related_work}
\subsection{Fully-supervised deep emotion recognition method based on peripheral physiological signals}
Deep learning-based methods have recently gained extensive attention due to their automatic abstract representation learning properties and have been shown to outperform machine learning methods in several studies ~\cite{DCNN, StressNAS, Lai_stress, CorrNet}. For example, Huynh et al. \cite{StressNAS} employed a neural architecture search, aiming to obtain the optimal architecture for emotion recognition among 10,000 manually designed deep neural networks for multimodal physiological signals. Lai et al. \cite{Lai_stress} proposed a residual temporal convolution-based deep neural network to capture the effective features of multimodal signals, resulting in state-of-the-art results for stress detection and emotion recognition tasks. 
The above results of deep learning-based approaches are encouraging for wearable emotion recognition. However, training a sufficiently accurate and generalizable model commonly depends on a large amount of labeled data, which is challenging for physiological data, as the 
annotation is time-consuming, expensive, and requires the intervention of domain experts. 
\subsection{Self-supervised learning (SSL) for limited labelled data}
\label{re_ssl}
To solve the overfitting problem introduced by the limited available data for supervised deep learning models, one common solution is data augmentation, i.e., applying different transformations on the original samples to obtain more abundant data. However, performing data augmentation could not introduce inter-subject variability during training \cite{sigrep}. 
Alternatively, a technique that does not require the intervention of labeled data is unsupervised representation learning, where a typical model is an autoencoder, which extracts meaningful representations through the compression and reconstruction of the unlabeled data.
Several studies have explored the feasibility of this technique for emotion recognition. 
In \cite{ross_unsupervised_2021}, stacked convolutional autoencoders were applied independently on unlabelled ECG and EDA data to obtain generalized latent representations for arousal classification, achieving better performance than the fully supervised approaches. Though this method effectively modeled the heterogeneity of multimodal signals, i.e. using different models to extract valid unimodal features, however, it neglected the collaborative and complementary nature of multimodality. Different from the previous approach, Zhang et al. \cite{CorrNet} presented a correlation-based emotion recognition algorithm (CorrNet), where intra-modal features are first obtained with separate convolutional autoencoders, followed by covariance and cross-covariance computation between each pair of modalities to obtain inter-modal features. 
 However, these unsupervised learning methods based on autoencoders did not introduce supervised signals in pre-training, thus may resulting in unsatisfactory performance.

Recently, a compelling branch in the field of unsupervised representation learning is self-supervised learning (SSL), which can effectively address the de-generalization issue posed by insufficient labeled data.
Unlike unsupervised learning which does not involve any labelled data, SSL is designed with a series of pretext tasks that introduce self-supervision to unlabelled data, enabling more effective representation learning for downstream tasks. 
Each unsupervised sample is automatically labeled through inherent dependencies and associations between the data without human intervention \cite{review_ssl}.  
The SSL model pre-trained on pseudo-labeled data is considered as powerful feature extractor for a variety of downstream tasks. 
In the domains of computer vision and natural language processing, SSL-based work such as SimCLR \cite{simclr}, Word2Vec \cite{word2vec}, and BERT \cite{bert} have exhibited competitive and even superior performance on a range of tasks. 
However, few studies have investigated the performance of SSL models on peripheral physiological signal data.
Sarkar et Etemad \cite{ecg_ssl} introduced a self-supervised representation learning framework for ECG-based emotion recognition, where the 1DCNN-based multi-task deep neural network 
is pre-trained with the objective of identifying the signal transformation types applied to unlabeled data.
Their study indicated that the pretext task based on transformation recognition can enable the model to better cope with potential variation factors in the data. However, not all time steps of a signal sequence are associated with the target event (i.e., a specific emotion). Thus, how to filter out irrelevant information during SSL for downstream tasks is an unsolved problem. Exploiting the synchrony of multimodal emotional responses is a potential solution. More specifically, multimodal physiological signals exhibit correlated or consistent temporal changes when emotions are elicited. In this way, modeling the correlation of multimodal signals in SSL can facilitate the capture of emotion-related components in unlabeled data. For multimodal emotion recognition, Dissanayake et al. \cite{sigrep} proposed a self-supervised contrastive learning approach, which aims to approximate the positive pairs while pushing the negative pairs away from each other. However, their SSL model is obtained by pretraining each modality independently, and thus again ignores the cross-modal correlations.
Therefore, more effective multimodal fusion strategies need to be developed for SSL-based wearable emotion recognition.
\subsection{Multimodal data fusion for emotion recognition}
Multimodal data fusion strategies can be generally categorized into: early fusion, intermediate fusion, and late fusion.
Most existing approaches for multimodal emotion recognition are based on early fusion, where multimodal data are combined as a whole before performing a learning task. Joint representations can be extracted directly from concatenated vectors with deep models such as 1DCNN\cite{1dcnn} and Bi-LSTM \cite{el_lstm_Zitouni}, which allow for encoding inter-modal correlations. 
However, since unimodal features are not learned explicitly (i.e., the heterogeneity of the multimodal signal is ignored), this fusion strategy is not effective in capturing intra-modal correlations.
Late fusion-based approaches \cite{DL_example1, Lai_stress} integrate the decisions of multiple independent learning models to predict emotion categories. Thus, in contrast to early fusion, this fusion approach ignores the connections and interactions between modalities.

Different from the previous fusion approaches, intermediate fusion enables both intra- and inter-modal correlation, where independent feature extractors are first applied to different modalities and the obtained unimodal features are then aggregated in an additional fusion module to further learn the joint representation. A variety of options exist for this fusion module.
For example, Shu and Wang \cite{IL_example1} adopt
ed the restricted Boltzmann machine (RBM) model to learn the joint probability distribution of multimodal low-level features to encode cross-modal information exchanges. 
Zhang et al. \cite{IL_example2} modeled the associations between multimodal features by introducing a regularization term to the objective function. 
More recently, the transformers have also gained popularity in intermediate fusion-based approaches ~\cite{VAT_transformer, tsai_multimodal_2019, sparse_trans} for video, audio and text.
Regarding studies on emotion recognition, 
Wu et al. \cite{VAT_transformer} proposed a multimodal Recursive Intermediate Layer Aggregation
(RILA) model, which was applied
between layers of unimodal deep transformers to capture interactions across modalities through the integration of multimodal intermediate representations. 
In this work, the transformers were employed to provide valid intermediate features. At the same time, they have also proved to be effective in merging multimodal data ~\cite{tsai_multimodal_2019, sparse_trans}. 
The attention mechanism can capture advanced patterns shared across modalities, thus exhibiting advantages over naive fusion strategies such as concatenation. 
In terms of practicality, multimodal emotion recognition based on the video, audio and text may not be
well suited to real-life scenarios, as it requires considerable
computational resources for long-term video stream analysis. In
contrast, wearable physiological signals can consistently predict emotions in a low-cost and objective way. 
However, the validity of transformer-based models has not been well established for wearable emotion recognition.
Meanwhile, video, audio and text-based approaches cannot be directly migrated to physiological data due to differences in data structures. In addition, they are susceptible to overfitting problems as they generally have a relatively deep architecture and follow a fully-supervised setup.
\section{Proposed Method}
\subsection{Overview}
Our goal is to employ unlabeled data for capturing generic representations of multimodal physiological signals in order to address the de-generalization problem introduced by a limited number of labeled samples. 
Hence, we propose a self-supervised learning (SSL) scheme using signal transformation recognition as a pretext objective. An illustration of the proposed approach is shown in Fig. \ref{Fig1:Overview}.  
In our work, three modalities measured by different sensors are considered: electrodermal activity (EDA), blood volume pressure (BVP) and skin temperature (TEMP). More formally, let $x_m \in\mathbb{R}^{N\times 1}$ represent a 1D time-domain signal from one of the M different modalities (in our work, M = 3), where N is the signal length. Given a set of \textit{n} transform functions $T = \{T_j(\cdot), j\in\{1, \dots, n\}\}$, the altered multimodal signal dataset can be generated by applying each transformation to individual modality. Based on this, one can easily build a pseudo-labeled dataset $\mathcal{L}= \{(T_j(x_m^i), y^i), y^i\in\{1, \dots, n\}, m\in\{1, \dots, M\}, i\in\interval{1}{|\mathcal{L}|}\}$ for unlabelled samples through self-supervision enabled by signal transformations. Then, the proposed model consisting of a multimodal encoder $E$ and modality-specific classifiers $C$ is pre-trained to predict the type of transformation applied to samples in $\mathcal{L}$. Ultimately, only the encoder part $E$ of the optimal model obtained after pre-training is retained and is expected to produce generalized multimodal representations in a variety of supervised downstream tasks. Details of the proposed SSL framework are as follows.

\subsection{Self-supervised learning of multimodal physiological signals}
\begin{figure*}[!htbp]
\centering
\includegraphics[scale = 0.5]{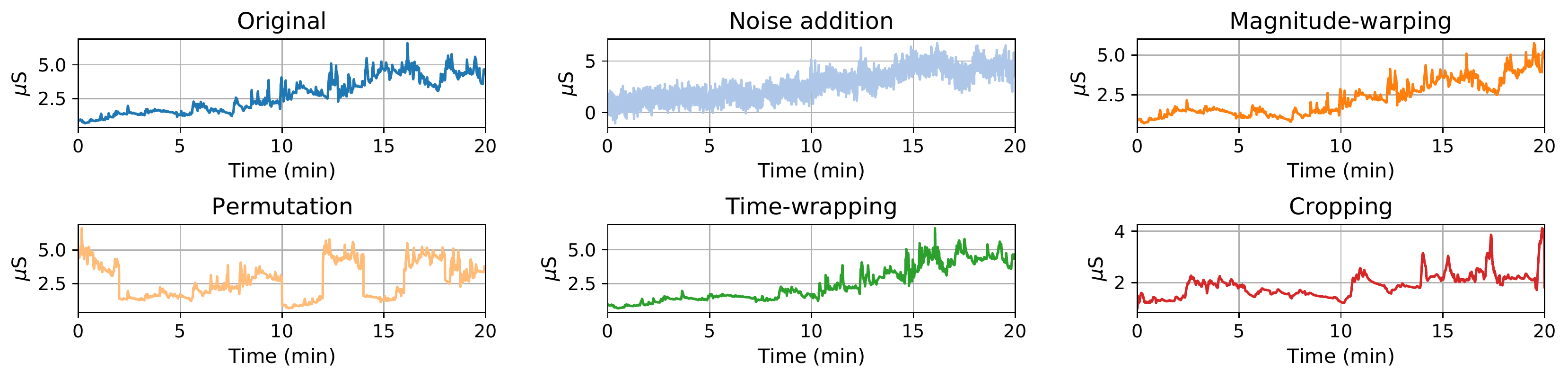}
\caption{The original signal and the disturbed signals after applying five transformations. For each modality, the raw signal data and the transformed signal data are stacked and fed into the proposed SSL model for multimodal representation learning.}
\label{Fig2:sig_trans}
\end{figure*}
\subsubsection{Pretext Task: signal transformation recognition}
\label{sec:sig_trans}
Signal transformation recognition was adopted as the pretext task in SSL, which proved to be effective in learning generalized representations for downstream tasks such as action recognition \cite{Saeed_activity_ssl} and emotion recognition \cite{ecg_ssl}.  
The random transformations used in the previous SSL methods are one of the common data augmentation techniques for time series, which can generally be classified into two categories: magnitude domain transformations and time domain transformations. The former interferes with the signal values while preserving the time step order, whereas the latter mainly affects the time scale.
Previous evaluations of SSL models based on individual transformation recognition 
 ~\cite{Saeed_activity_ssl, ecg_ssl} have indicated that \textit{Noise addition} and \textit{Scaling} ranked highly for magnitude domain transformations, while \textit{Permutation} and \textit{Time-warping} performed outstandingly well among time domain transformations.
Meanwhile, according to the review of time series augmentation strategies \cite{wearable_augment, data_aug_review_1}, though most of the suggested transformations have been adopted in previous SSL-based work, two transformations have not been thoroughly evaluated: \textit{Magnitude-warping} and \textit{Cropping}.
Ultimately, we selected the five transformations: \textit{Permutation}, \textit{Time-warping}, \textit{Noise addition}, \textit{Magnitude-warping} and \textit{Cropping} for the pretext task. The reason why \textit{Scaling} was omitted is that \textit{Magnitude-warping} can be seen as a special variant of \textit{Scaling}\footnote{\textit{Scaling} multiplies time series values by
a random scalar whereas \textit{Magnitude-warping} distorts the signal values by a smooth curve.}. 

The above signal transformations are performed on all three modalities and the resulting transformed signal data is fed into the proposed SSL model as input along with the original multimodal signal data. 
Fig. \ref{Fig2:sig_trans} shows the effect of these deformations on a sample of the EDA signal. Details of each transformation are described in subsequent paragraphs. Here, for simplicity, we write the above-mentioned 1D signal $x_m$ uniformly as $x(t)$, where t represents the time step.\\
\textbf{Magnitude domain transformations}:
\begin{itemize}
    \item \textbf{Gaussian noise addition:} The original signal $x(t)$ is disturbed by white Gaussian noise $z(t)$, which can be extracted from a zero-mean normal distribution ${\mathcal  {N}}(0,\sigma ^{2})$. By assigning a preferred signal-to-noise ratio (SNR), the variance $\sigma ^{2}$ (i.e., the average power of the noise) of the distribution ${\mathcal {N}}$ can be derived from the following formula $10^{(P_{sig}-SNR)/10}$, where ${P_{sig}}$ is the average power of the signal. In the end, the noised signal is calculated as $x(t) + z(t)$.
    \item \textbf{Magnitude-warping:}
    The magnitudes of the original signal are altered by a random smooth curve formed by cubic spline interpolation function $\phi(\cdot)$. In the end, the transformed signal can be calculated as $x(t)\cdot \phi(x(t))$.
\end{itemize}
\textbf{Time domain transformations}:
\begin{itemize}
    \item \textbf{Permutation:} The original signal is split into \textit{n} non-overlapping segments $x(t)=\{x_1, x_2, ..., x_n\}$, which are then temporally disrupted and eventually recombined together to form the permuted signal $x(t)=\{x_{p1}, x_{p2}, ..., x_{pn}\}$, where $\{p1, p2, ..., pn\}$ is a shuffled version of the original order.
    \item \textbf{Time-warping:} The original signal is divided into \textit{n} non-overlapping segments $x(t)=\{x_1, x_2, ..., x_n\}$, half of which are randomly selected to be stretched by a linear interpolation function $F(x_i, k)$, where $k$ is the stretch factor, and the remaining half of the segments are squeezed by the function $F(x_i, 1/k)$, where $1/k$ is the squeeze factor. The time-warped signal can be concatenated from the transformed segments and finally resized to the original length. 
    \item \textbf{Cropping:} The original signal is divided into \textit{n} non-overlapping segments $x(t)=\{x_1, x_2, ..., x_n\}$, one of which is randomly selected and resampled to the original length.
\end{itemize}
By identifying the signal transform types, our model is expected to learn a more robust and generalized representation against disturbances in the magnitude or time domains. For example, \textbf{Magnitude-warping} and \textbf{Gaussian noise addition} can simulate different types of real-world noise, such as measurement errors, signal artefacts caused by the subject's body movements, etc. For time-domain transformations, \textbf{Permutation} perturbs the order of time steps to prompt the model for capturing time-domain dependencies between data points, \textbf{Time-warping} simulates duration variations in emotional responses by stretching or squeezing time steps, and \textbf{Cropping} allows the model to be more robust to changes in the temporal location of emotional events.
\subsubsection{Self-supervised multimodal representation learning network architecture}
The proposed SSL multimodal deep neural network consists of two key elements, namely the encoder $E$ and the modality-specific transformation classifiers $C$. The encoder E can be further subdivided into temporal convolution-based modality-specific encoders $E_p$ and transformer-based shared encoder $E_s$, where $E_p$ models the heterogeneity of multimodal signals and $E_s$ activates cross-modal information exchange. Ultimately, the multimodal features obtained from the encoder are used as input to $C$ for identifying transformation types for each modality. The implementation of these key components is described in the following paragraphs.\\
\textbf{Modality-specific encoder:} 
Considering the heterogeneity of the multimodal signals, separate encoders are first employed for each modality, with a temporal convolution-based network acting as the backbone to capture low-level intra-modal correlation information.
The temporal convolutional network (TCN) \cite{Bai2018AnEE}, in a nutshell, is a combination of dilated causal convolution and residual connections, with parallel computational capability and robust gradients at optimization, thus demonstrating better performance than traditional recurrent networks, such as LSTM and GRU. One basic TCN consists of several residual blocks. The most central components of each 
block are two dilated causal convolution layers. 
The causality can be easily achieved when the output at the current moment $t$ depends only on the elements of the past historical moments up to $t$ in the previous layer. Meanwhile, the dilation operation injects holes in the standard convolution map, thereby increasing the reception field. 
More formally, given the transformed 1D signal of modality $m$: $x_m^{\prime}=T_j(x_m) \in\mathbb{R}^{N\times 1}$ with $N$ time steps, and a filter $f$ of size k, the dilated convolution on time step $t$ can be defined as
\begin{equation}
\begin{aligned}
F(t) = \sum_{i=0}^{k-1} f(i)\cdot x_m^{\prime}(t-d\cdot i)
\end{aligned}
\label{eq:3}
\end{equation}
where d is the dilation factor. 
Following each convolutional layer is a weight normalization layer for the convolution filter, a rectified linear unit (ReLU) layer and a dropout layer for regularization. In the end, a residual connection is created between the input and output of the block, where a $1\times1$ convolution is introduced to eliminate the mismatch in channel numbers between the input and output. 
\begin{figure*}[!htb]
\centering
\includegraphics[scale = 1.2]{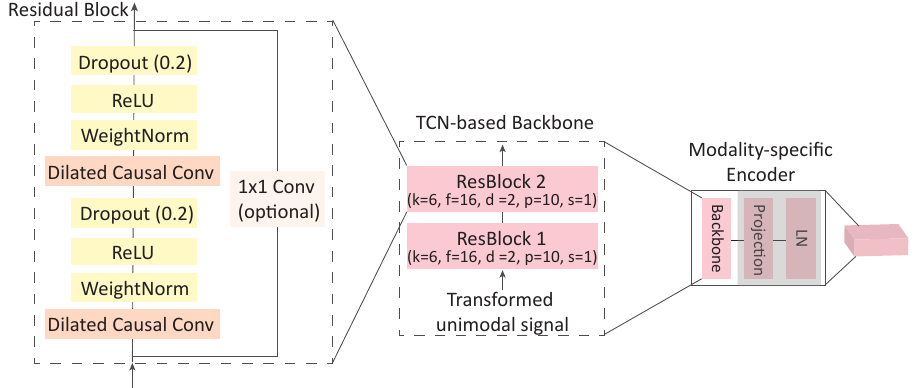}
\caption{Modality-specific backbone based on temporal convolutional network (TCN). Each backbone consists of two residual blocks for capturing low-level features for transformed unimodal signals $x_m^{\prime}$. (k: kernel size, f: number of filters, d: dilation factor, p: padding size, s: stride size, weightnorm: weight normalization for convolution filters)}
\label{Fig3:tcn_backbone}
\end{figure*}
Fig.3 illustrates the detailed structure of the TCN-based backbone.
The dilated causal convolution layers in two residual blocks are equipped with 16 filters with a kernel size of 6, where the dilation factors are 1 and 2, respectively. Zero-padding of 5 and 10 are also introduced to ensure that the input and output sequences are of the same length. Subsequently, a modality-specific projection head (i.e., a linear fully connected layer with 128 units) and a layer normalization are then applied to map the low-level features to a higher dimensional embedding space. Finally, the output of the modality-specific encoder $E_p$ is:\\
\begin{equation}
\begin{aligned}
z_m = LayerNorm(MLP(TCN(x_m^{\prime}))) \in \mathbb{R}^{N\times d}
\end{aligned}
\label{eq:4}
\end{equation}
where d is the embedding dimension.\\
\textbf{Shared encoder:} 
As mentioned in Section \ref{re_ssl}, encoding of the coordination and interaction between multimodal signals is essential in order to learn generic representations related to the downstream emotion recognition tasks. 
This can be done through the transformer in which each modality identifies components of other modalities that are highly correlated with itself through the attention mechanism for better signal transformation classification.
To achieve this, the low-level features $z_m$ of each modality are first stacked to form a multimodal embedding $z_{multi} = [z_1, \dots, z_m, \dots, z_M]\in \mathbb{R}^{MN\times d}$. The scaled dot-product attention proposed in \cite{NIPS2017_3f5ee243} is then applied to calculate the dependencies between different modalities:
\begin{equation}
\begin{aligned}
Attn(Q, K, V) = softmax(\frac{QK^T}{\sqrt{d}})V
\end{aligned}
\label{eq:5}
\end{equation}
where $Q$, $K$, $V$ represent queries, keys and values, respectively. More intuitively, the attention layer acts as a weighted sum of values $V$, where the attention weight associated with each value is generated by the compatibility of the query with its corresponding key. For our shared encoder $E_s$, queries, keys and values are derived through a linear mapping of multimodal features $z_{multi}$, and the resulting output of the attention layer is:
\begin{equation}
\begin{aligned}
z_{multi}^a = Attn(z_{multi}W^Q, z_{multi}W^K, z_{multi}W^V) 
\end{aligned}
\label{eq:6}
\end{equation}
where $W^Q$, $W^K$, $W^V\in \mathbb{R}^{d\times d}$ are the projection matrices. 
\begin{figure*}[!htb]
\centering
\includegraphics[scale = 1.1]{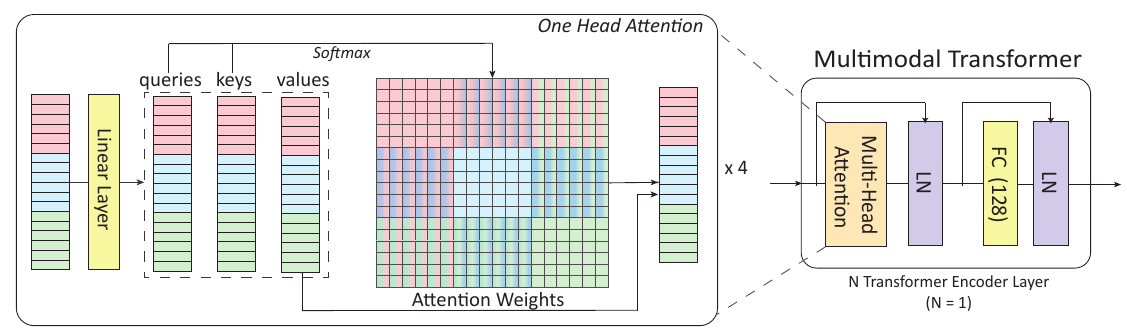}
\caption{Shared encoder based on the multimodal transformer.
(FC: fully-connected layer with 128 units, LN: layer normalization)}
\label{Fig4:transformer}
\end{figure*}
Fig. \ref{Fig4:transformer} presents the process of generating attention weights from multimodal embeddings, where cross-modal communications are activated. For our shared encoder, the one-layer vanilla transformer block proposed in \cite{NIPS2017_3f5ee243} with four-head attention is implemented. The feedforward layer dimension is set to 128. ReLU is selected as the activation function for intermediate layers and a rate of 0.2 is used for Dropout operation. In addition, we did not introduce positional coding information for the stacked multimodal inputs. Since the features of each modality are generated by different encoders, the network performance may not benefit from positional encoding in the context of heterogeneous input. This is further explored in the ablation study (Section \ref{ablation_component}).\\
\textbf{Modality-specific classification head:} 
The multimodal features $h_{multi}\in \mathbb{R}^{MN\times d}$ extracted from the shared encoder $E_s$ are then decomposed to $[h_1,\dots,h_m,\dots,h_M]$ for identifying the type of signal transformation applied to each modality.
\begin{figure}[!htb]
\centering
\includegraphics[scale = 1.1]{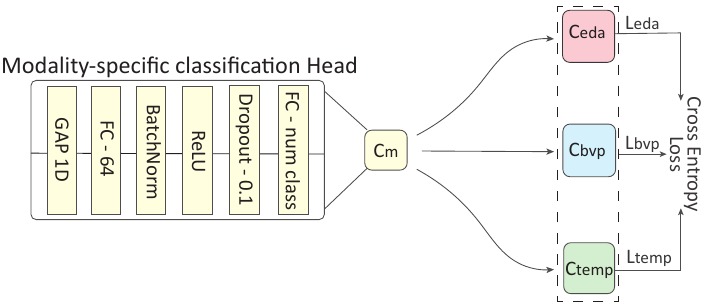}
\caption{Modality-specific classification head $C_m$ for signal transformation recognition task. 
(GAP: 1D global average pooling, FC: fully-connected layer, BatchNorm: batch normalization, num class: number of signal transformations, i.e., 6 in our work.)}
\label{Fig5:ssl_classification_head}
\end{figure}
A modality-specific classification head $C_m$ is shown in Fig. \ref{Fig5:ssl_classification_head}. 1D global average pooling is first applied across all time steps of unimodal features, followed by a fully-connected layer with 64 units. 1D Batch Normalization is placed before the ReLU layer for more efficient learning and a Dropout layer with a rate of 0.1 is applied to avoid over-fitting. The final fully-connected layer is equipped with a softmax activation function, where the unit number is determined by the number of signal transformations $n$ (i.e., $n=6$ in our work, 5 transformations plus the original version). In the end, the proposed model is optimized on the pseudo-labeled dataset $\mathcal{L}$
through the total loss $L_{total}$ which is a combination of cross-entropy losses of individual modalities (i.e., EDA, BVP, TEMP in our work):
\begin{equation}
 L_{m}  = - \frac{1}{|{\mathcal{L}}|} \sum_{i=1}^{|{\mathcal{L}}|} y^{i} log(C_m(h_m^{i}))
\label{eq:7}
\end{equation}
\begin{equation}
\begin{aligned}
 L_{total} & = \sum_{i=1}^M L_m = L_{eda} + L_{bvp} + L_{temp}
\end{aligned}
\label{eq:8}
\end{equation} 
\subsection{Multimodal emotion recognition based on physiological signals}
After pre-training the proposed SSL model with the pretext task on unlabelled data, only the encoder part $E$ is reserved for extracting efficient multimodal representations in a variety of supervised downstream tasks. In this work, we select emotion recognition as our downstream task. A classification head $C_{emo}$ is applied to the output of the encoder $E$ to generate class probabilities for labeled samples $\mathcal{L}_{sup} =  \{(x_m^i, y^i), , y^i\in\{1, \dots, e\}, m\in\{1, \dots, M\}\}$, where $e$ is the number of the emotion classes. The emotion classification head is constructed in the same way as $C_m$, except that it accepts multimodal features from encoder $E$. After the multimodal transformer, features from each modality are first passed through the 1D global average pooling layer, then the flattened unimodal features are concatenated and processed successively through a fully-connected layer with 192 hidden units, a Batch Normalization layer, ReLU activation function, a Dropout layer with a rate of 0.2 and a second fully-connected layer with the number of hidden units equal to the number of emotion classes for prediction.
Finally, the proposed model is optimized through the minimization of cross entropy loss $L_{sup}$.
\begin{equation}
 L_{sup}  = - \frac{1}{|{\mathcal{L}_{sup}}|} \sum_{i=1}^{|{\mathcal{L}_{sup}}|} y^{i} log(C_{emo}(E(x^{i})))
\label{eq:9}
\end{equation}
\section{Datasets}
\subsection{PRESAGE Dataset}
\begin{figure*}[!htb]
\centering
\includegraphics[scale = 0.7]{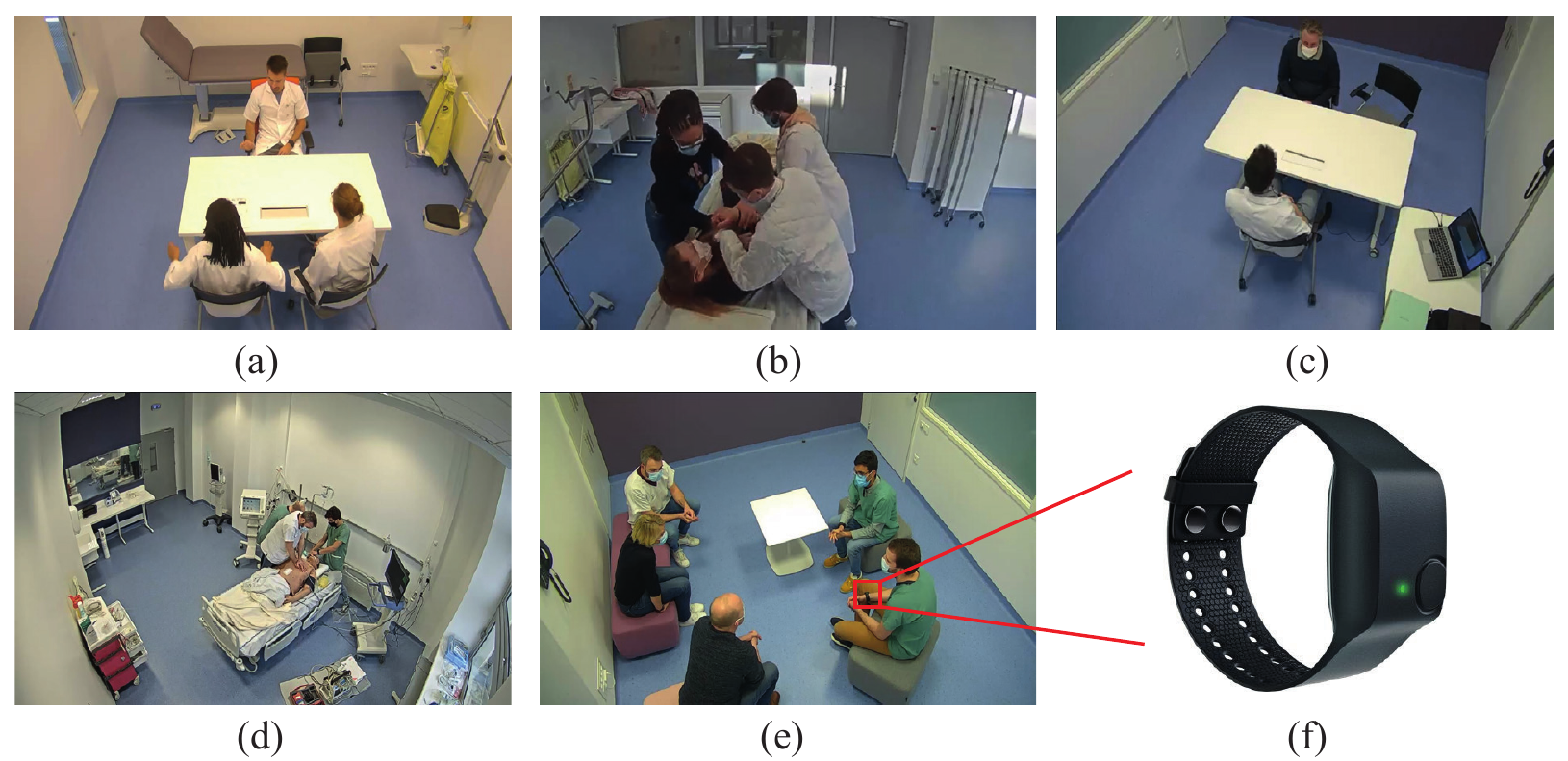}
\caption{Images of different scenarios captured by cameras placed in the simulation training room: (a): Doctor consultation, (b): Prevention of escape for patients in an acute agitated state, (c): Second consultation for patients with suicidal tendencies, (d): Management of cardiac arrest/severe head injury/chest trauma, (e): Diagnostic announcement and (f): the wearable sensor Empatica E4 wristband  used for physiological data collection during the simulation training.}
\label{Fig6:presage_description}
\end{figure*}

The \textbf{PRESAGE dataset} is a large-scale multimodal physiological signal dataset for emotion analysis. The data acquisition is done at the Presage training center\footnote{\url{https://medecine.univ-lille.fr/presage}.} in Lille, France, whose mission is to ensure the training of medical students and health professionals through immersion in a recreated hospital environment, where the high-tech mannequins or hired actors, take the place of the patients and students act as doctors. In order to analyze the students' emotional state during the simulation training to optimize the educational program, a large amount of unlabeled multimodal physiological data has been collected from 201 trainees (104 males and 97 females) during five different medical simulation scenarios.
Fig. \ref{Fig6:presage_description} (a-e) shows the images of different scenarios captured by the cameras installed in the simulation room. 
The data collection protocol was approved by the Institutional Review Board of University of Lille with the reference number 2022-626-S108 and
all trainees were given a consent form prior to training and were required to fully read the form and provide a signature. 
To allow students to perform normal medical simulation activities under interference-free conditions, \textit{Empatica E4 Wristband} (Fig. \ref{Fig6:presage_description} (f)), an invasive wearable biometric sensor was adopted to continuously record multimodal physiological signal data of high quality with different frequencies: 3-axis Accelerometer (ACC, 32Hz), Blood Volume Pressure (BVP, 64Hz), Electrodermal Activity (EDA, 4Hz), Skin Temperature (TEMP, 4Hz), Heart Rate (HR, 1Hz), Inter-beat Interval (IBI). In this work, we employed data from
three modalities: EDA, BVP and TEMP collected in the five scenarios for self-supervised multimodal representation learning.
\subsection{WESAD Dataset}
The \textbf{WESAD dataset} \cite{schmidt_introducing_2018} is a multimodal dataset for stress and emotion recognition.
Following a study protocol in a restricted laboratory setting, three affective states, namely baseline, stress and amusement, were elicited from 15 subjects during which physiological and motion signals were collected by two separate sensors: RespiBAN (chest-worn device) and Empatica E4 (wrist-worn device). Since we focus on wearable affective computing, only blood volume pressure (BVP, 64 Hz), electrodermal activity (EDA, 4 Hz) and temperature (TEMP, 4 Hz) captured by Empatica E4 were applied to the classification task. According to previous work~\cite{schmidt_introducing_2018,StressNAS,Lai_stress}, a stress detection task (\textit{non-stress vs stress}) and a emotion recognition task (\textit{baseline vs stress vs amusement}) can be performed on the WESAD dataset for supervised learning, where the non-stress class is a combination of the baseline and amusement classes.
\subsection{CASE Dataset}
The \textbf{CASE dataset} \cite{case_dataset_2019} is a multimodal emotion recognition dataset with continuous annotations. Eight video clips were employed to stimulate four different emotions: amusing, boring, relaxing and scary from 30 subjects. During the experiment, subjects were required to self-assess their own emotional experiences using an annotation interface based on valence-arousal scores, while six physiological signals were recorded at a frequency of 1000 Hz. In our work, we selected blood volume pressure (BVP), electrodermal activity (EDA) and skin temperature (TEMP) signals as in the self-supervised dataset for the classification task. We adopted the same approach as in the literature ~\cite{sigrep,CorrNet} for the mapping from continuous values of valence and arousal to discrete classes, resulting in a binary (\textit{low vs high valence/arousal}) and a three-class (\textit{low vs medium vs high valence/arousal}) classification problem for supervised learning.
\subsection{K-EmoCon Dataset}
The \textbf{K-EmoCon dataset} \cite{kemocon} is a multimodal dataset with multiperspective annotations for emotion recognition in social interactions. 32 subjects were divided into 16 groups for a two-person debate, during which facial expressions, upper body posture, audio signals, EEG signals and peripheral physiological signals were recorded by different sensors. In our experiments, only blood volume pressure (BVP), electrodermal activity (EDA) and skin temperature (TEMP) signals measured by Empatica E4 were retained for downstream emotion recognition tasks. Tripartite 
annotations, i.e., self-annotations, partner annotations and external observer annotations were employed to assess subjects' affective states during the debate. Based on the previous work \cite{sigrep}, we categorized the arousal- and valence-based annotations into discrete classes, thus forming a binary (low vs high valence/arousal) and a three-class (low vs
medium vs high valence/arousal) classification problem for supervised learning. 
\section{Experiments and Results}
\subsection{Data Preprocessing}
To eliminate artifacts, we first applied a low-pass Butterworth filter with a cutoff frequency of 0.5 Hz for the EDA and TEMP signals, while the same type of filter with a cutoff frequency of 2 Hz is selected for the BVP signal in PRESAGE, WESAD and K-EmoCon dataset. For the CASE dataset, a low-pass filter with a cutoff frequency of 2 Hz was utilized to clean these three signals. Moreover, we performed z-score normalization as in \cite{ssl_ecg} for each signal recording to reduce the variation in physiological responses between different subjects. Since the four datasets involved in the experiments were collected using sensors with different sampling frequencies, we then uniformly downsampled all signals in the different datasets to the most frequently occurring frequency, i.e., 4 Hz. Subsequently, based on previous work \cite{StressNAS, Lai_stress}, we segmented the signal recordings of all datasets into windows of length 60 s with 99.5\% overlap for PRESAGE and WESAD, 99\% and 95\% overlap for CASE and K-EmoCon, respectively. If the data in a window corresponds to multiple 
labels, we adopt the same strategy as in the previous work \cite{sigrep}, i.e., choosing the one with the majority as the final label. Table \ref{table:data_prepro} concludes the learning tasks corresponding to each dataset and the number of samples created after data segmentation. The last column in the table lists the total size of each dataset, where the first dimension represents the total number of samples, while the second and third dimensions represent the signal length at a frequency of 4 Hz in a 60 s window after segmentation (i.e., 240) and the number of modalities (i.e., three modalities: BVP, EDA and TEMP), respectively.
\begin{table*}[h]
\caption{The learning tasks assigned to each dataset and the corresponding distribution of samples between classes in different datasets. (P: Pretext task, D: Downstream task.)}
\centering
\resizebox{15cm}{!}{
\begin{tabular}{c||c||c||c||c}
\hline
Dataset & Type & Task & Category (no. of samples) & Total Size\\\hline
PRESAGE & P & \makecell{Transformation \\ Recognition} & \makecell{Original version and\\ five transformations (681641)}& (4089846, 240, 3)\\\hline
\multirow{ 2}{*}{WESAD} & \multirow{ 2}{*}{D}  & Stress-2  & stress (36279), non-stress (85574)& \multirow{2}{*}{(12185, 240, 3)} \\
&  & Emotion-3  & baseline (66859), stress (36279), amusement (18715) & \\\hline
\multirow{ 4}{*}{CASE} & \multirow{ 4}{*}{D} & Arousal-2  & low (33211), high (61919) & \multirow{4}{*}{(95130, 240, 3)}\\
& & Valence-2  & negative(32017), positive (63113)  \\
& & Arousal-3  & low (4847), medium (26898), high (63385) \\
& & Valence-3  & negative(9312), neutral (56870), positive (28948) \\\hline
\multirow{ 4}{*}{K-EmoCon} & \multirow{ 4}{*}{D} & Arousal-2  & low (3729), high (1488) & \multirow{4}{*}{(5217, 240, 3)} \\
& & Valence-2  & negative(4050), positive (1167) & \\
& & Arousal-3  & low (1783), medium (1904), high (1530) &  \\
& & Valence-3  & negative(1783), neutral (1904), positive (1530) &  \\
\hline
\end{tabular}
}
\label{table:data_prepro} 
\end{table*}
\subsection{Implementation and model training}
The training process of our SSL-based approach consists of two main phases. The first phase is to pre-train the proposed model on the PRESAGE dataset using automatically generated pseudo-labels for signal transformation identification. A set of transformation parameter vectors (15, 10, 9, 1.05, 4, 0.2) was chosen based on the experimental results of the previous study \cite{ssl_ecg} as SNR, magnitude warping variance coefficient, number of permutation segments, number of time warping segments, time-warping stretching coefficient, and number of cropping segments for each modality to generate the five transformations mentioned in Section \ref{sec:sig_trans}. The pre-training process of the proposed model took approximately 26 hours on an NVIDIA RTX 6000 GPU. 
The second phase retains only the encoder part of the pre-trained model to extract valid, generalized representations for emotion recognition on WESAD, CASE and K-EmoCon datasets. We did not introduce these three public datasets into pre-training stage in order to verify the knowledge transfer ability of the learned features across different datasets. Ultimately, the proposed model was installed using Pytorch. 
The optimal models for the pretext and downstream tasks were obtained by the SGD (Stochastic Gradient Descent) optimizer with weight decay parameter of 5e-7 to avoid overfitting. For the first phase (self-supervised pre-training), learning rate, batch size and the number of epochs are set to 5e-3, 32 and 20, respectively. For the second phase (supervised emotion recognition), the learning rate, batch size and number of epochs are set to 1e-4, 128, 20 on WESAD dataset, while for CASE and K-EmoCon datasets, these parameters were set to 1e-3, 64 and 64, respectively.
\subsection{Evaluation metric and protocol}
For a fair comparison, we adopted the same experimental protocol as in ~\cite{StressNAS, Lai_stress, schmidt_introducing_2018, CorrNet, sigrep}, i.e. Leave-One-Subject-Out cross validation, which has the benefit of examining the generalization ability of the model to unpresented subject data. Two metrics, accuracy and F1-score applied in ~\cite{StressNAS, Lai_stress, schmidt_introducing_2018, CorrNet, sigrep} were selected to evaluate the performance of the proposed approach on the emotion recognition task. Accuracy represents the proportion of correctly classified samples to the total number of samples. F1-score is considered as 
a harmonic mean of the precision and recall, which is suggested for evaluating imbalanced datasets. 
\subsection{Baseline Models}
\label{sec:model_comp}
Since the exploration of wearable emotion recognition based on peripheral physiological signals has not been well established, a series of baseline models based on fully-supervised learning, unsupervised learning, and self-supervised learning were implemented in addition to available state-of-the-art methods to provide a more comprehensive and reliable performance comparison.
The followings are brief descriptions of these models:
\vskip 0.1in
\noindent 
Supervised learning-based methods:
\begin{itemize}
  \item \textbf{SimpDCNN} \cite{Saeed_activity_ssl}: it is a simple convolutional network consisting of three convolutional blocks with kernel sizes of 24, 16 and 8, each followed by a ReLU activation and a dropout layer.
  \item \textbf{MulT} \cite{MulT}: it is a transformer-based multimodal fusion method applied to video, audio and text. The unimodal data is first passed through a temporal convolutional network to obtain low-level features, then transformers based on cross-modal attention and self-attention mechanisms are applied successively for effective fusion.
  \item \textbf{ResNet} \cite{resnet_org}: it is a 1D convolution-based residual network adapted to physiological signals proposed in ~\cite{resnet_18_1, resnet_18_2}, which is constructed similarly to ResNet-18, consists mainly of 8 residual blocks with batch normalization (BN) operation and ReLU activation function, where each block contains two convolutional layers. The three modalities: BVP, EDA, TEMP are fed into this network as multi-channel signals.
  \item \textbf{Ours (Supervised)}: it is our proposed multimodal network, trained in a fully-supervised manner.
 \end{itemize}
In addition, three additional supervised methods were applied for the performance comparison on the CASE and K-EmoCon datasets since they lacked baseline results compared to the WESAD dataset.
\begin{itemize}
  \item \textbf{DCNN} \cite{1dcnn}: it employs a four-layer 1D convolutional neural network to extract modality-specific features, and a three-layer fully connected network connected at the bottom of the network for classification.
  \item \textbf{Attn-BiLSTM} \cite{attnbilstm}: it applies a multilayer bidirectional LSTM for capturing valid temporal information for multimodal signals. The attention mechanism was applied to select the most relevant multimodal representation of the emotional state as input for a fully connected layer-based classifier.
  \item \textbf{MMResLSTM} \cite{mmreslstm}: it uses separate four-layer LSTM-based models for multimodal signals with residual connections. Moreover, the weights of the LSTM layers of both modalities are shared to activate cross-modal communication. 
 \end{itemize}
Unsupervised learning-based methods:
\begin{itemize}
  \item \textbf{Autoencoder}: it is an autoencoder with the same encoder part as our proposed model, while the decoder part consists of three transposed convolutional blocks for the reconstruction of the BVP, EDA, TEMP signals. Each unimodal decoder consists of four-layer transposed convolution with the same parameters as the convolutional layers in the encoder.
\end{itemize}
Self-supervised learning-based methods:
\begin{itemize}
  \item \textbf{SigRep} \cite{sigrep}: it adopts a similar  model architecture to SimCLR \cite{simclr}, containing an encoder of four inception-inspired blocks and a projection head consisting of fully connected layers, where each inception block consists of 1D convolutional layers with different kernel sizes and a maximum pooling layer in parallel. The model is applied independently to each signal modality for contrastive representation learning.
  \item \textbf{BENDR} \cite{BENDR}: it is a simpler version of wav2vec 2.0 \cite{wave2vec} that was applied to EEG signals. We adapted it for application to peripheral physiological signals at low frequencies. The multi-channel signal consisting of BVP, EDA, TEMP is first passed through a four-layer convolution with kernel sizes of 3, 2, 2, 2, where the GeLU is chosen as the activation function along with GroupNorm and Dropout operations, and the obtained low-level features are randomly masked and fed to the same transformer as our proposed model. The final output features are used to reconstruct the masked features.
\end{itemize}
For a fair comparison, we used the code provided by the authors of the above methods and applied the same experimental setup. If the code is not available, we followed the parameters provided in these works for the model implementation. For those models initially designed for non-peripheral physiological signals, the parameters have been slightly adjusted to match the low-frequency wearable data for proper operation.
\subsection{Experimental Results}
\begin{table}[!htbp]
\caption{Performance comparison of different emotion recognition tasks with state-of-the-art methods on the WESAD dataset. (SL: supervised learning methods, UL: unsupervised learning methods, SSL: self-supervised learning methods, S: supervised, F: frozen, T: fine-tuned.)
}
\centering
\resizebox{8.5cm}{!}{
\begin{tabular}{|c|l|c|c|c|c|}
\hline
\multirow{ 2}{*}{Type} & \multirow{ 2}{*}{Methods} &\multicolumn{2}{c|}{Stress-2} & \multicolumn{2}{c|}{Emotion-3}\\\cline{3-6}
 &  & Acc & F1 &  Acc & F1 \\\hline
 \multirow{8}{*}{SL} 
& LDA\cite{schmidt_introducing_2018} & 86.46  & 83.77 & 68.85 & 58.18 \\
& RF\cite{schmidt_introducing_2018} & 88.33 & 86.10 & 76.17 & 66.33\\
& SimpDCNN\cite{Saeed_activity_ssl} & 90.12 & 88.22 & 78.30 & 74.59\\
& MulT\cite{MulT} & 91.76 & 91.17 & 81.09 & 78.27\\
& ResNet\cite{resnet_org} & 91.93 & 90.97 & 80.85 & 79.63\\
& StressNAS\cite{StressNAS} & 92.87  & - & 81.78 & - \\
& Res-TCN\cite{Lai_stress} & 94.16 & 93.62 & 83.69 & 81.61 \\
& Ours (S) & 93.83 & 92.55 & 84.81 & \textbf{83.76}\\
\hline
UL & Autoencoder & 91.51 & 90.33 & 80.39 & 79.13\\\hline
\multirow{6}{*}{SSL} & SigRep\cite{sigrep} (F) & 92.71 & 91.99 & 81.11 & 78.92\\
& SigRep\cite{sigrep} (T) & 94.91 & 93.09 & 84.27 & 82.35\\\cline{2-6}
& BENDR\cite{BENDR} (F) & 92.53 & 91.72 & 81.98 & 79.71\\
& BENDR\cite{BENDR} (T) & 93.19 & 92.55 & 82.44 & 80.69\\\cline{2-6}
& Ours (F) & 94.81 & 93.69 & 83.81 & 82.01\\
& \textbf{Ours (T)} & \textbf{96.29} & \textbf{95.11} & \textbf{84.94} & 82.60\\
\hline
\end{tabular}
}
\label{table:wesad} 
\end{table}
\subsubsection{Comparision with state-of-the-art methods}
Emotion-related classification tasks were performed on WESAD, CASE, K-EmoCon datasets to evaluate the performance of the proposed SSL model. Tables \ref{table:wesad}, \ref{table:case}, \ref{table:kemocon} summarize performance comparisons with the state-of-the-art fully supervised, unsupervised, and self-supervised learning-based methods. 
For the SSL-based approaches, we report the results under two training modes: \textbf{Frozen} (F) and \textbf{Fine-Tuned} (T).
The first mode refers to freezing the pre-trained encoder part and updating only the parameters of the classification head in the downstream classification tasks, which is designed to investigate the effectiveness of the learned self-supervised multimodal features. 
The second mode employs the pre-trained encoder parameters for model initialization and updates all parameters normally to examine the performance gain relative to the \textbf{Frozen} mode. 
From the tables, first, it can be observed that our fully-supervised model obtained better performance than other supervised learning approaches in most emotion recognition tasks, confirming the effectiveness of the proposed architecture.
Secondly, regarding our SSL model, the comparison results indicated that, under the \textbf{Frozen} mode, our method achieved superior performance over other fully supervised, unsupervised, and self-supervised based approaches on 6 out of 10 tasks, demonstrating the generalization and high discrimination of the representation learned through the SSL pretext task. In addition, the performance of our model was improved in the \textbf{Fine-Tuned} mode, further narrowing the gap with supervised baselines and thus achieving state-of-the-art results in 8 out of 10 tasks. Additionally, it is interesting to note that as the number of supervised samples decreases from WESAD to CASE to K-EmoCon, the higher the performance gain obtained by our SSL-based approach with respect to the supervised approaches. This can be attributed to the fact that supervised learning methods are more prone to overfitting than self-supervised learning methods on low data regimes. Further research on the performance comparison of these two types of methods on limited data is presented in Section \ref{litte_data_comp}. 
Thirdly, in comparison with non-supervised learning methods, we significantly improved the performance of SigRep and BENDR, especially on the CASE and K-EmoCon datasets. The source of this performance gap may be related to the deployed fusion strategies, in addition to the selected pretext tasks. 
SigRep \cite{sigrep} learned effective representations for each modality independently through contrastive learning, whereas BENDR \cite{BENDR} regarded multimodal signals as a whole to reconstruct obscured multimodal features. Thus, these two approaches ignored the encoding of inter- and intra-modal correlations, respectively. The impact of different SSL fusion strategies on downstream performance is later investigated in Section \ref{ablation_fusion_strategy}. Furthermore, the results of the Autoencoder are inferior to other SSL methods. This may be due to the unsupervised nature of its pre-training process which results in more redundant patterns being captured that are irrelevant to the downstream tasks.
\begin{table*}[!htbp]
\caption{Performance comparison of different emotion recognition tasks with state-of-the-art methods on the CASE dataset. (SL: supervised learning methods, UL: unsupervised learning methods, SSL: self-supervised learning methods, S: supervised, F: frozen, T: fine-tuned.)
}
\centering
\resizebox{13cm}{!}{
\begin{tabular}{|c|l|c|c|c|c|c|c|c|c|}
\hline
\multirow{ 2}{*}{Type} & \multirow{ 2}{*}{Methods} &\multicolumn{2}{c|}{Valence-2} & \multicolumn{2}{c|}{Valence-3}
& \multicolumn{2}{c|}{Arousal-2}& \multicolumn{2}{c|}{Arousal-3}  \\
\cline{3-10}
 &  & Acc & F1&  Acc & F1
& Acc & F1 & Acc & F1\\\hline
\multirow{ 7}{*}{SL} 
& SimpDCNN\cite{Saeed_activity_ssl} & 71.33 & 68.74 & 59.20 & 51.95 & 67.16 & 61.60 & 56.80 & 53.85\\
& DCNN\cite{1dcnn} & 72.35 & 69.96 & 59.78 & 52.80 & 69.63 & 63.43 & 56.09 & 53.51\\
& MMResLSTM\cite{mmreslstm} & 73.34 & 70.96 & 60.78 & 53.09 & 71.12 & 68.06 & 57.41 & 54.69 \\
& Attn-BiLSTM\cite{attnbilstm} & 74.25 & 71.27 & 61.97 & 53.64 & 70.40 & 66.52 & 58.27 & 54.76\\
& MulT\cite{MulT} & 74.81 & 73.17 & 63.14 & 62.50 & 71.28 & 70.44 & 62.15 & 58.48 \\
& ResNet\cite{resnet_org} & 75.29 & 74.62 & 62.89 & 62.18 &  72.35 & 72.19 & 65.46 & 59.69\\
& Ours (S) & 76.94 & 75.06 & 64.58 & 63.29 & 74.15 & 72.86 & \textbf{66.32} & \textbf{61.78}\\
\hline
\multirow{ 2}{*}{UL} & Autoencoder & 73.23 & 72.05 & 60.77 & 57.32 &  69.16 & 67.13 & 60.08 & 55.12\\
& CorrNet\cite{CorrNet} & 76.37 & 76.00 & 60.15 & 53.00 & 74.03 & 72.00 & 58.22 & 55.00 \\\hline
\multirow{ 6}{*}{SSL}
& SigRep\cite{sigrep} (F) & 71.74  & 64.78 & 63.85 & 54.97  & 70.79 & 67.28 & 63.09 & 56.99  \\
& SigRep\cite{sigrep} (T) & 73.29  & 69.84 & 64.63 & 55.68  & 72.08 & 69.45 & 64.88 & 58.81  \\\cline{2-10}
& BENDR\cite{BENDR} (F) & 72.94 & 68.48 & 61.56 & 50.86 &  72.04 & 67.43 & 62.37 & 55.63\\
& BENDR\cite{BENDR} (T) & 72.33 & 67.62 & 62.15 & 53.03 & 71.51 & 67.32 & 63.52 & 57.01\\\cline{2-10}
& Ours (F) & 77.49 & 75.85  &  65.51  & 64.07  & 73.67  & 70.76  & 65.09 & 59.64 \\
& \textbf{Ours (T)} & \textbf{78.57} & \textbf{77.74}  &  \textbf{66.64}  & \textbf{64.85}  & \textbf{74.98}  & \textbf{73.10} & 66.19 & 60.56\\
\hline
\end{tabular}}
\label{table:case} 
\end{table*}
\begin{table*}[!htbp]
\caption{Performance comparison of different emotion recognition tasks with state-of-the-art methods on the K-EmoCon dataset. (SL: supervised learning methods, UL: unsupervised learning methods, SSL: self-supervised learning methods, S: supervised, F: frozen, T: fine-tuned.)
}
\centering
\resizebox{13cm}{!}{
\begin{tabular}{|c|l|c|c|c|c|c|c|c|c|}
\hline 
\multirow{ 2}{*}{Type} & \multirow{ 2}{*}{Methods} &\multicolumn{2}{c|}{Valence-2} & \multicolumn{2}{c|}{Valence-3}
& \multicolumn{2}{c|}{Arousal-2}& \multicolumn{2}{c|}{Arousal-3}  \\
\cline{3-10}
 &  & Acc & F1&  Acc & F1
& Acc & F1 & Acc & F1\\\hline
\multirow{ 7}{*}{SL} 
& SimpDCNN\cite{Saeed_activity_ssl} & 77.14 & 70.06 & 59.67 & 48.98 & 72.48 & 61.21 & 46.49 & 38.34\\
& DCNN\cite{1dcnn} & 78.72 & 72.09 & 61.97 & 51.39 & 73.67 & 65.53 & 49.91 & 39.24\\
& Attn-BiLSTM\cite{attnbilstm} & 79.76 & 72.19 & 62.56 & 54.35 & 73.30 & 66.23 & 46.95 & 46.77\\
& MMResLSTM\cite{mmreslstm} & 78.79 & 72.76 & 61.25 & 51.65 & 74.31 & 67.88 & 44.68 & 37.19 \\
& MulT\cite{MulT} & 80.13 & 76.72 & 63.95 & 59.07 &  74.19 & 72.49 & 49.25 & 47.86 \\
& ResNet\cite{resnet_org} & 80.53 & 78.04 & 64.60 & 62.22 &  74.35 & 73.20 & 50.09 & 46.77\\
& Ours (S) & 81.51 & 78.60 & 64.07 & 60.83 & 75.17 & 73.62 & 50.42 & 47.52\\\hline
UL & Autoencoder & 80.58 & 75.58 & 63.65 & 58.32 &  71.56 & 69.10 & 48.83 & 47.10\\\hline
\multirow{ 6}{*}{SSL} & SigRep\cite{sigrep} (F) & 78.98 & 73.15 & 63.00 & 54.07 & 73.36 & 67.80 & 47.85 & 42.58 \\
& SigRep\cite{sigrep} (T) & 79.14 & 73.55 & 61.74 & 52.69 & 73.94 & 68.71 & 48.56 & 43.90 \\\cline{2-10}
& BENDR\cite{BENDR} (F) & 79.83 & 72.62 & 61.38 & 53.20 &  72.86 & 66.16 & 50.68 & 48.03\\
& BENDR\cite{BENDR} (T) & 78.73 & 72.15 & 61.85 & 54.47 &  73.82 & 69.46 & 52.88 & 51.24\\\cline{2-10}
& Ours (F) & 82.95 & 80.07 & 66.97 & 61.28 & 74.79 & 73.40 & 50.76 & 48.66 \\
& \textbf{Ours (T)} & \textbf{84.14} & \textbf{81.08} & \textbf{68.37} & \textbf{63.10} & \textbf{76.40} & \textbf{74.29} & \textbf{54.60} & \textbf{52.34} \\
\hline
\end{tabular}}
\label{table:kemocon} 
\end{table*}
\subsubsection{Self-supervised learning vs Supervised learning on limited labeled data}
\label{litte_data_comp}
\begin{figure*}[!htb]
\centering
\includegraphics[scale = 0.5]{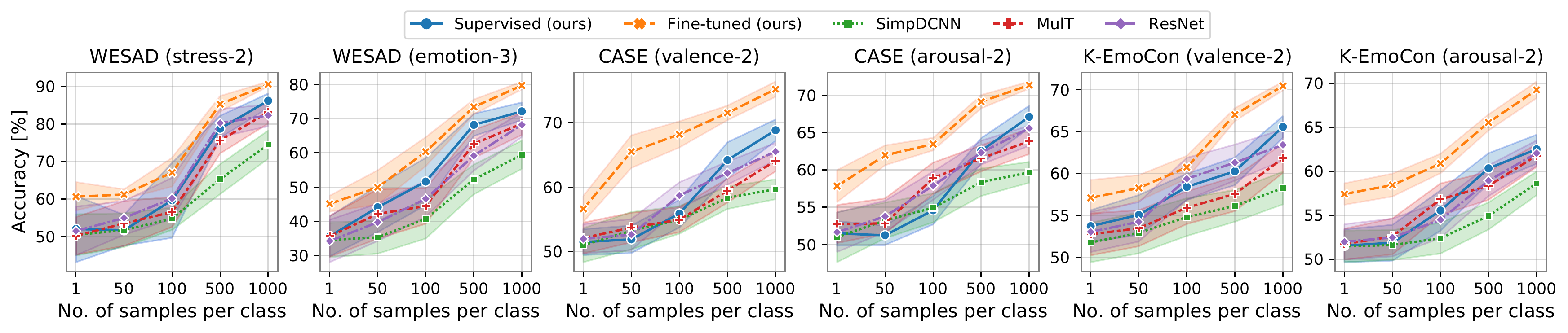}
\caption{Performance comparison with state-of-the-art supervised learning-based methods on limited labeled data sampled from the three emotion recognition datasets. The horizontal axis of each subplot is the number of randomly selected samples from each class, varying from 1 to 1000, while the vertical axis is the corresponding average accuracy.}
\label{Fig:little_data}
\end{figure*}

In the previous section, our self-supervised approach presented state-of-the-art performance on emotion recognition tasks with all labeled data in the dataset. 
To further investigate the effectiveness of our fine-tuned model on a limited number of labeled samples, we performed a comparison with four supervised learning models: our proposed model with fully-supervised learning, MulT \cite{MulT}, ResNet \cite{resnet_org} and SimpDCNN \cite{Saeed_activity_ssl}. 
MulT and ResNet were selected since they share similar structures to our model and are the best-performing supervised models in addition to ours. Besides, SimpDCNN, as a low-complexity model, is not prone to overfitting on limited data, allowing for a more comprehensive performance comparison.
We implemented a similar sampling procedure reported in \cite{Banville_2021, jiang_eeg}, i.e., 1, 50, 100, 500, and 1000 samples were randomly selected for each class in the three datasets for training the classification model. This process was executed 50 times independently for different numbers of samples. The resulting average accuracy and the corresponding standard deviation of all compared models are illustrated in Fig. \ref{Fig:little_data}. 
First, our fine-tuned model consistently outperforms other supervised learning-based models for sample sizes varying from 1 to 1000 on the emotion recognition tasks of all three datasets. Among supervised learning-based methods, SimpDCNN exhibited the poorest results, over which our SSL model could achieve significant performance gains of 6.84\% - 21.19\% for different downstream tasks. 
Our fully-supervised model yields the highest results compared to other supervised models, whereas the fine-tuned model initialized by self-supervised learning parameters continues to enhance performance by 5.24\% - 13.63\%. 
Second, for all downstream tasks, the standard deviation obtained by our fine-tuned model is narrower with respect to the supervised learning-based deep models, demonstrating its superior generalization ability across different samples.
The above findings are consistent with those reported in \cite{Newell_2020_CVPR} that the advantage of the self-supervised learning-based method is its better regularisation on low data regimes to avoid overfitting problems compared to fully-supervised methods. As the amount of available labeled data increases, the difference in performance between the two types of models gradually decreases. Overall, the comparison results suggest that the proposed method can produce more meaningful and robust representations for wearable emotion recognition than fully-supervised methods, offering a potential solution to the problem of little labeled data.
\begin{figure}[htbp]
\centering
\includegraphics[scale = 0.98]{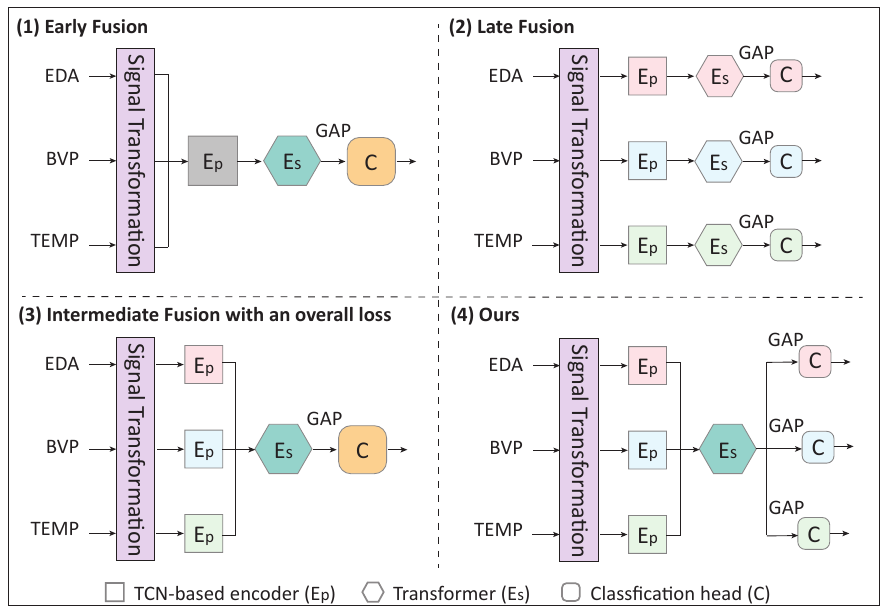}
\caption{Different architectures used in the ablation studies of fusion strategies. (GAP: 1D global average pooling applied before classification.)}
\label{Fig:ablation_fusion}
\end{figure}
\subsection{Ablation Studies}
Different types of ablation experiments were designed and conducted on the WESAD, CASE, and K-EmoCon datasets to verify the validity of the proposed method. 
The encoder part of the models involved was trained in freezing mode and the obtained emotion recognition results are reported in the following sections.
\subsubsection{Ablation study of different fusion strategies}
\label{ablation_fusion_strategy}
To demonstrate the effectiveness of the selected fusion strategy, we conducted ablation studies on different SSL fusion strategies.
Three variants of the proposed model based on early fusion, late fusion, and intermediate fusion strategies were implemented for comparison. The corresponding model architectures used for comparison are illustrated in Fig \ref{Fig:ablation_fusion}. In all variant models, the TCN-based encoder $E_p$, transformer $E_s$ and signal transformation classification head $C$ all share the same parameter settings as the proposed model (details are given in Section 3.2.2). 
For the \textbf{Early fusion} setup, we treated the multimodal physiological signal as a whole, i.e. a multichannel signal, from which multimodal representations will be learned directly.
For the \textbf{Late fusion} setup, separate encoders were applied to individual modalities to extract unimodal features for classification. 
In addition, the third variant model has the same fusion strategy as ours, where unimodal features were first captured and then concatenated to learn more advanced multimodal features. The difference, however, is that this model performs classification by multimodal features. This is to verify the necessity of conducting modality-specific classification in the proposed method, and we refer to this setup as \textbf{Intermediate fusion with an overall loss}. Consequently, the corresponding evaluation results are listed in Table \ref{table:ablation_fusion}. Our model consistently achieved the best performance on all datasets, demonstrating the effectiveness of the selected fusion strategy, i.e., intermediate fusion. In addition, the intermediate fusion-based models performed better than those based on the other two fusions. This can be attributed to the fact that the intermediate fusion simultaneously models the heterogeneity and coordination of multimodal physiological signals, whereas the other two fusion approaches only consider one of these two properties. Furthermore, the third setting \textbf{Intermediate fusion with an overall loss} performs slightly worse than our model, affirming the importance of modality-specific classification. The benefit of applying modality-specific loss functions is that it forces the model to learn, for each modality, generic features that are robust to perturbations in the time or magnitude domain, while the application of an overall loss fails to distinguish each modality's contribution to the learned representation.
\begin{table*}[h]
\caption{Ablation study of different fusion strategies: average accuracy and F1-score obtained for emotion recognition on WESAD, CASE, and K-EmoCon dataset using different variant models. 
(S-2: Stress-2, E-3: Emotion-3, V-2: Valence-2, A-2: Arousal-2, Inter w/ ol: Intermediate fusion with an overall loss.)
}
\centering
\begin{tabular}{|c|c|c|c|c|c|c|c|c|c|c|c|c|}
\hline 
\multirow{ 3}{*}{Type} & \multicolumn{4}{c|}{WESAD} & \multicolumn{4}{c|}{CASE} & \multicolumn{4}{c|}{K-EmoCon} \\
\cline{2-13}
 & \multicolumn{2}{c|}{S-2} & \multicolumn{2}{c|}{E-3} &  \multicolumn{2}{c|}{V-2} & \multicolumn{2}{c|}{A-2}
&\multicolumn{2}{c|}{V-2}& \multicolumn{2}{c|}{A-2}\\\cline{2-13}
 &  Acc & F1 & Acc & F1
& Acc& F1& Acc & F1 & Acc& F1 & Acc & F1\\\hline
Early  & 91.22 & 89.94 & 80.82 & 78.79 & 73.01 & 72.20 & 69.13 & 67.06 & 79.20 & 74.12 & 71.57 & 69.60 \\
Late & 93.02 & 91.73 & 81.48 & 80.91 & 75.58 & 72.27 & 71.96 & 68.52 & 80.94 & 76.43 & 72.81 & 70.87  \\
Inter w/ ol & 93.53 & 92.77 & 82.82 & 81.62 & 76.69 & 73.52 & 72.24 & 69.11 & 81.48 & 77.22 & 73.06 & 71.29 \\
\hline
\textbf{Ours}  & \textbf{94.81} & \textbf{93.69} & \textbf{83.81} & \textbf{82.01} & \textbf{77.49} & \textbf{75.58} & \textbf{73.67} & \textbf{70.76} & \textbf{82.95} & \textbf{80.07} & \textbf{74.79} & \textbf{73.40 } \\\hline
\end{tabular}
\label{table:ablation_fusion} 
\end{table*}
\begin{table*} [htp]
\caption{Ablation study of different modalities and their combinations: average accuracy and F1-score obtained with different modality combinations in the downstream emotion recognition tasks, where the best performing individual modality and bimodal combinations for each task are underlined. 
(S-2: Stress-2, E-3: Emotion-3, V-2: Valence-2, A-2: Arousal-2.)
}
\centering
\resizebox{12.5cm}{!}{
{\Large
\begin{tabular}{|c|cc|cc|cc|cc|cc|cc|}
\hline 
\multirow{ 3}{*}{Modality} & \multicolumn{4}{c|}{WESAD} & \multicolumn{4}{c|}{CASE} & \multicolumn{4}{c|}{K-EmoCon} \\
\cline{2-13}
 & \multicolumn{2}{c|}{S-2} & \multicolumn{2}{c|}{E-3} &  \multicolumn{2}{c|}{V-2} & \multicolumn{2}{c|}{A-2}
&\multicolumn{2}{c|}{V-2}& \multicolumn{2}{c|}{A-2}\\\cline{2-13}
 &  Acc & F1 & Acc & F1
& Acc & F1 & Acc & F1 & Acc & F1 & Acc & F1\\\hline
EDA  &  \underline{92.36} & \underline{90.58} & \underline{78.72} & \underline{75.90} & 75.21  & 74.80 & \underline{72.15} & \underline{70.13} & 80.65 & 74.60 & \underline{73.09} & \underline{72.69}\\
BVP &  87.82 & 86.35 & 75.20 & 68.80 & \underline{75.90} & \underline{75.15} & 69.23 & 65.07 & \underline{80.76} & \underline{74.13} & 72.88 &  70.67\\
TEMP & 78.15 & 76.91 & 69.86 & 65.37 & 71.64 & 68.66 & 68.97 & 62.16 & 79.02 & 72.78 & 72.52 & 70.42 \\\hline
EDA + BVP  & \underline{93.73} & \underline{92.38} & \underline{82.32} & \underline{80.61} & \underline{76.26} & \underline{75.13} & 72.13 & 70.27 & 80.87 & 75.48 & 73.14 & 71.78\\
EDA + TEMP &  90.95 & 89.62 & 79.74 & 76.09 & 76.03 & 74.97 & \underline{72.92} & \underline{70.54} & \underline{81.70} & \underline{77.77} & \underline{74.61} & \underline{72.93}\\
BVP + TEMP & 84.82 & 80.45 & 72.88 & 66.16 & 72.35 & 71.31 & 71.14 & 68.04 & 80.12 & 75.05 & 72.43 & 69.86 \\\hline
All & \textbf{94.81} & \textbf{93.69} & \textbf{83.81} & \textbf{82.01} &  \textbf{77.49} & \textbf{75.85} & \textbf{73.67} & \textbf{70.76} & \textbf{82.95}  & \textbf{80.07} & \textbf{74.79} & \textbf{73.40}\\
\hline
\end{tabular}}
\label{table:ablation_modality_combination} 
}
\end{table*}
\subsubsection{Ablation study of different modalities}
\label{ablation_signal_modality}
We conducted an ablation study of three modalities: EDA, BVP, TEMP and their combinations to explore their performance on emotion recognition tasks. The resulting average accuracies and F1-scores of our model are summarized in Table \ref{table:ablation_modality_combination}. First, for the unimodal performance, the EDA signal performs outstandingly well among all the modalities, especially when detecting stress and arousal states. This is consistent with previous findings that EDA is one of the most relevant indicators of stress \cite{Healey_stress} and has even been adopted as ground truth in some studies ~\cite{Wijsman_stress, Hernandez_stress} for the stress analysis of other signals. In addition, it has been proven to correlate linearly with arousal \cite{Hernandez_stress}. 
In the bimodal-based classification, we first observed that the \textbf{BVP+EDA} setup performed better on the stress-related tasks (i.e. S-2 and E-3 on the WESAD dataset) than the other setups. 
This suggests that the BVP signal and the EDA signal are highly coordinated and correlated when the stress state is elicited, making their combination more effective for detection. This finding is quite reasonable. The BVP signal contains information on heart rate (HR) and heart rate variability (HRV) thus providing a strong correlation with stress states. In \cite{ALBERDI201649}, HRV and EDA were identified as the most relevant physiological indicators for the real-time stress detection task. 
Secondly, the \textbf{EDA+TEMP} setup achieved the best performance on the classification task regarding arousal level. This finding is supported by previous research \cite{SATO2020107974} which indicated that EDA and TEMP had a positive and negative correlation with arousal scores respectively. Lastly, our model achieved performance gains on both bimodal and trimodal data in most cases, confirming again its effectiveness for multimodal fusion.
\subsubsection{Ablation study of missing modalities}
\begin{figure*}[!htb]
\centering
\includegraphics[scale = 0.5]{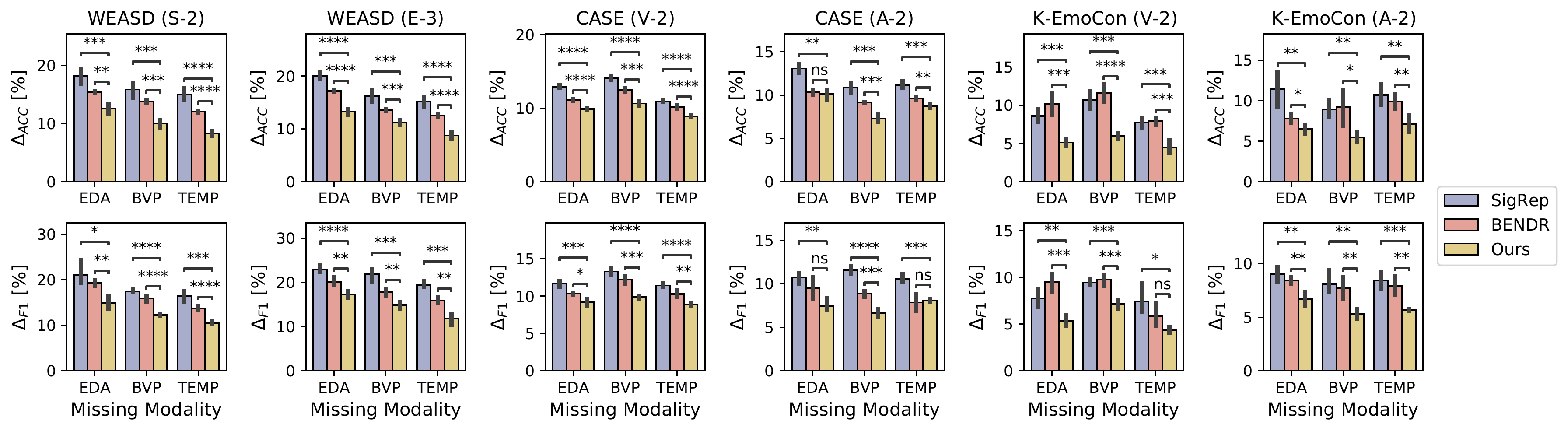}
\caption{Evaluation results of the robustness of the SSL methods in the presence of missing modalities. The horizontal axis of each subplot represents the name of the missing modality, while the vertical axis represents the drops in model performance compared to the case of complete modalities, where the metrics of the vertical axes in the first and second rows are accuracy and F1-score, respectively. (ns: no significant difference; $\ast$: $p<0.05$, the more asterisks, the more significant the difference.)}
\label{Fig:missing_modalities}
\end{figure*}

We investigate the robustness of the proposed SSL model when a signal modality is missing in downstream tasks, which is quite common in real-world scenarios. There exist a variety of experimental setups for incomplete modalities. Based on \cite{Ma_missing_modality}, we selected the most challenging one, i.e., a modality is missing in both training and testing of the downstream task, where 50\% of the multimodal samples were first randomly selected and subsequently the data values of a specific modality were set to 0 to simulate its absence. The robustness of the proposed SSL model was measured by calculating its difference in performance in two cases: one with all modalities present and one with missing modalities. The smaller the difference, the more robust the model is considered to be. The above experimental procedure was repeated 10 times. Additionally, we benchmarked our model against the SSL baseline models: SigRep and BENDR. Fig. \ref{Fig:missing_modalities} presents the average degradation in accuracy and F1-score of the compared models when a modality is missing in different downstream tasks. A series of t-tests were further conducted on the performance differences for a more systematic robustness comparison. From the evaluation results, we can first observe that the performance drops of our model are significantly lower ($p < 0.05$) than other SSL models on most tasks. This demonstrates the superiority of the proposed method in terms of robustness. Second, we also note that the impact of missing modalities on the robustness of SSL methods is task-dependent. For downstream tasks related to stress and arousal levels, more severe performance declines could be obtained in the absence of the EDA signal, compared to the other two modalities. This result indicates the importance of the EDA signal for identifying these two emotional states. Similarly, missing the TEMP signal also leads to a considerable reduced performance in arousal-based recognition, whereas, in the valence-based tasks, the loss of the BVP signal has the greatest impact on performance. The above results, consistent with those in Section \ref{ablation_signal_modality}, reconfirm the effect of different modalities on specific emotion recognition.
\begin{table*}[h]
\caption{Ablation study of different model components: average accuracy and F1-score obtained for emotion recognition on WESAD, CASE, and K-EmoCon dataset using different variant models. (S-2: Stress-2, E-3: Emotion-3, V-2: Valence-2, A-2: Arousal-2)}
\centering
\resizebox{15cm}{!}{
\begin{tabular}{|c|c|c|c|c|c|c|c|c|c|c|c|c|}
\hline 
\multirow{ 3}{*}{Model Variants} & \multicolumn{4}{c|}{WESAD} & \multicolumn{4}{c|}{CASE} & \multicolumn{4}{c|}{K-EmoCon} \\
\cline{2-13}
 & \multicolumn{2}{c|}{S-2} & \multicolumn{2}{c|}{E-3} &  \multicolumn{2}{c|}{V-2} & \multicolumn{2}{c|}{A-2}
&\multicolumn{2}{c|}{V-2}& \multicolumn{2}{c|}{A-2}\\\cline{2-13}
 &  Acc & F1 & Acc & F1
& Acc& F1& Acc & F1 & Acc& F1 & Acc & F1\\\hline
No TCN  & 91.09 & 90.42 & 79.86 & 78.77 & 64.15 & 62.20 & 66.87 & 57.87 & 79.39 & 75.34 & 71.24 & 66.59   \\
No Transformer & 92.18 & 91.55 & 81.08 & 79.30 & 74.65 & 71.53 & 70.04 & 68.40 & 80.67 & 76.17 & 72.09 & 70.66 \\\hline
With fixed PE  & 93.49 & 91.63 & 82.38 & 80.47 & 76.37 & 74.45 & 72.59 & 70.05 & 80.32 & 77.43 & 73.35 & 71.20\\
With learnable PE & 92.68 & 91.32 & 82.42 & 81.24 & 76.46 & 75.35 & 73.16 & 70.33 & 81.59 & 78.62 & 74.08 & 72.09\\\hline
\textbf{Our Model}  & \textbf{94.81} & \textbf{93.69} & \textbf{83.81} & \textbf{82.01} & \textbf{77.49} & \textbf{75.58} & \textbf{73.67} & \textbf{70.76} & \textbf{82.95} & \textbf{80.07} & \textbf{74.79} & \textbf{73.40 } \\\hline
\end{tabular}}
\label{table:ablation_model} 
\end{table*}
\begin{table*}[!htbp]
\caption{Ablation study of individual signal transformations and their combinations: average accuracy and F1score obtained for emotion recognition on WESAD, CASE and K-EmoCon datasets using different transformations in self-supervised pertaining, where the best-performing transformations and combinations of transformations in each task are underlined. (N: Noise addition, M: Magnitude-warping, P: Permutation, T: Time-warping, C: Cropping.)
}
\centering
\resizebox{15cm}{!}{
\begin{tabular}{|c|cc|cc|cc|cc|cc|cc|}
\hline 
\multirow{ 2}{*}{Type} & \multicolumn{4}{c}{WESAD} & \multicolumn{4}{c}{CASE} & \multicolumn{4}{c|}{K-EmoCon} \\
\cline{2-13}
 & \multicolumn{2}{c|}{S-2} & \multicolumn{2}{c|}{E-3} &  \multicolumn{2}{c|}{V-2} & \multicolumn{2}{c|}{A-2}
&\multicolumn{2}{c|}{V-2}& \multicolumn{2}{c|}{A-2}\\\cline{1-13}
\textbf{Single} & Acc & F1 & Acc & F1 & Acc& F1& Acc & F1 & Acc& F1 & Acc & F1\\\cline{1-13}
N & 90.18 & 89.16 & 78.04 & 76.24 & 73.53 & 70.42 &  69.16 & 61.48 & 79.69 & 74.30 & 70.23 & 67.38 \\
M  & 89.74 & 87.86 & 76.90 & 73.77 & 68.75 & 67.46 &  68.28 & 66.94 & 80.13 & 75.44 & 71.11 & 70.54 \\
P  & 91.20 & 89.33 & 78.38 & 75.67 & \underline{73.61} & \underline{71.17} & \underline{71.46} & \underline{69.82} & 80.22 & 78.12 & \underline{71.69} & \underline{70.82}\\
T  & \underline{91.34} & \underline{90.87} & \underline{81.15} & \underline{80.44} & 71.17 & 69.53 & 70.81 & 69.01 & \underline{80.39} & \underline{78.31} & 70.59 & 67.90\\
C  & 89.48 & 88.06 & 79.21 & 77.38 & 69.35 & 62.46 & 69.17 & 66.60 & 79.68 & 74.07 & 70.87 & 69.05\\\hline
\textbf{Same Domain} & Acc & F1 & Acc & F1 & Acc& F1& Acc & F1 & Acc& F1 & Acc & F1\\\hline
N+M  & 89.69  & 88.43 & 77.52 & 75.59  &  \underline{75.87} & \underline{74.22}  & \underline{71.83}  &  \underline{69.62} &  \underline{81.46} & \underline{78.95}  &  \underline{73.14} &  \underline{72.08} \\
P+T+C  &  \underline{93.67} & \underline{92.88}  & \underline{82.31}  & \underline{80.47}  &  73.71 &  71.95 & 70.92  & 69.05  &  80.77 & 78.58  &  72.66 &  71.11 \\[0.5pt]\hline
\textbf{Cross Domain} & Acc & F1 & Acc & F1 & Acc& F1& Acc & F1 & Acc& F1 & Acc & F1\\\hline
N+P & 92.15 & 91.12  & 80.02 & 78.93  & \underline{75.77} & \underline{73.57} & \underline{72.85} & \underline{70.09} & 80.26  & 78.54  &  73.09 & 71.73  \\
M+P &  91.75 & 90.47  & 79.24  & 77.31  &  74.01 &  72.92 &  72.39 & 69.87  &  80.34 &  79.15 &  71.89 & 70.76  \\
N+T & \underline{92.95} & \underline{91.69}  & \underline{83.11} & \underline{81.14} & 73.51  &  71.21 &  70.03 &  68.91 & 81.57  &  79.43 &  72.04 & 69.75  \\
M+T & 91.51 & 90.18 & 79.85 & 77.52 &  71.45 & 70.07  & 71.61  & 69.14  &  80.29 &  79.04 & 71.39  & 69.61 \\
N+C & 91.28 & 90.15 & 78.81 & 76.66 & 73.96  & 72.60  & 70.57  & 68.44  &  79.52 &  77.83 &  73.13 &  71.52\\
M+C & 90.08 & 89.62  & 78.27  & 76.40 & 71.90  & 69.45  & 69.53 & 67.89  &  \underline{82.14} & \underline{79.61} &  \underline{73.89} &  \underline{72.51}\\[0.5pt]\hline
\textbf{All} & \textbf{94.81} & \textbf{93.69} & \textbf{83.81} & \textbf{82.01} & \textbf{77.49} & \textbf{75.85} & \textbf{73.67} & \textbf{70.76} & \textbf{82.95} & \textbf{80.07} & \textbf{74.79} &\textbf{73.40}\\
\hline
\end{tabular}}
\label{table:ablation_transformation} 
\end{table*}
\subsubsection{Ablation study of different model components}
\label{ablation_component}
We also investigate the impact of different model components on the performance of downstream classification tasks. To validate the contributions of the modality-specific encoder and the shared encoder, we designed two alternative models: \textbf{No TCN} and \textbf{No Transformer}.
\textbf{No TCN} eliminates the temporal convolution network (TCN) where unimodal data is passed directly through the projection layer (i.e. a fully connected layer with 128 units) in the modality-specific encoder shown in Fig. \ref{Fig3:tcn_backbone} and the resulting unimodal low-level features are then concatenated as a whole and fed into the transformer.
\textbf{No Transformer} removes the multimodal transformer, where unimodal features are first extracted by modality-specific encoders and then averaged along the time dimension by the 1D global average pooling (illustrated in Fig. \ref{Fig5:ssl_classification_head}) for the final classification tasks.
Table \ref{table:ablation_model} present the classification results of the above two variant models on the three datasets. 
Our proposed model enhances both the performance of \textbf{No TCN} and \textbf{No Transformer} models on all classification tasks across different datasets, highlighting the importance of capturing the heterogeneity and cross-modal correlation of multimodal signals simultaneously.
Subsequently, we examined whether the addition of positional encoding could lead to better performance for the transformers with heterogeneous embedding as input. We employed two types of positional encoding (PE): \textbf{With fixed PE} and \textbf{With learnable PE} in the transformer and compared their performance with our PE-free model. \textbf{With fixed PE} added the fixed positional encoding obtained from sine and cosine functions of different frequencies as proposed in \cite{NIPS2017_3f5ee243} to the input embedding of the multimodal transformer while \textbf{With learnable PE} adopted the same learnable positional encoding in \cite{learnable_pe}.
Table \ref{table:ablation_model} also show the classification results of the proposed model with different PE setting.
We observed that temporal context information injected by two types of PE did not contribute to model performance on all classification tasks as expected. This can be attributed to the fact that the multimodal embeddings generated by the separate encoders already own different structures, hence the additional positional information introduces redundancy into the model.
\subsubsection{Ablation study of different signal transformations}
We further explored the impact of using individual transformations and their combinations in the pretext task on downstream emotion recognition performance.  
As mentioned in Section \ref{sec:sig_trans}, the five transforms employed can be divided into two classes, i.e., magnitude domain transformations and time domain transformations.
Therefore, the types of combinations are arranged accordingly as combinations of transformations within the same domain and combinations of transformations across domains. The evaluation results obtained on different emotion classification tasks are presented in Table \ref{table:ablation_transformation}.
First, we noticed that \textbf{Permutation} and \textbf{Time-Warping}, which perturbed the temporal order and duration of events within the window, performed best among the individual signal transformations, which is consistent with the results in ~\cite{Saeed_activity_ssl, ecg_ssl}, demonstrating the necessity to encode the temporal relationships of signals for emotion recognition.
Second, the pre-trained models obtained by combining the same domain or cross-domain transformations generally perform better than those based on individual transformations. The performance of these combinations varies depending on the specific task. For the same domain transformation combinations, \textbf{P+T+C} performs better for stress-related tasks, whereas \textbf{N+M} is more appropriate for arousal and valence-based tasks. For the cross-domain combinations, \textbf{N+T} exhibited the best performance on the classification tasks regarding stress, while \textbf{N+P} and \textbf{M+C} performed best in predicting the arousal and valence states.
Finally, we found that models based on cross-domain combinations outperformed those based on the same domain combinations in two-thirds of the downstream tasks. Meanwhile, our pre-trained models using the full set of transformations consistently achieved superior performance in the classification tasks. This can be attributed to the fact that different types of transformations inject diverse prior knowledge for multimodal representation learning, thus contributing to the generalizability of the network.
\section{Conclusion}
In this paper, we have proposed a self-supervised multimodal representation learning framework for wearable emotion recognition. Signal transformation recognition is defined as a pretext task, where a large amount of unsupervised data is automatically labeled by the imposed signal transformation category for pre-training of the SSL model. Subsequently, the encoder part of the pre-trained model consisting of a temporal convolution network and transformer is maintained to extract effective multimodal representations for the downstream task, i.e. emotion recognition. Eventually, we executed the pre-training on a large-scale unrestricted 
emotion dataset PRESAGE and verified the validity of the proposed method on three public multimodal emotion recognition datasets. Experimental results indicated that our approach surpassed fully-supervised, unsupervised, and self-supervised learning methods, achieving state-of-the-art results in various emotion-related tasks. 
Additionally, the proposed method performs better than the fully-supervised learning approach on limited labeled data, demonstrating its superior generalization ability to avoid overfitting problems. A series of ablation studies also confirmed the efficiency of the designed model architecture.


%



\ifCLASSOPTIONcompsoc
  \section*{Acknowledgments}
\else
  \section*{Acknowledgment}
\fi

The proposed work was supported by the French State, managed by the National Agency for Research (ANR) under the Investments for the future program with reference ANR-16-IDEX-0004 ULNE. 

\ifCLASSOPTIONcaptionsoff
  \newpage
\fi

\vspace{-15 mm}
\begin{IEEEbiography}
[{\includegraphics[width=1in,height=1.25in]{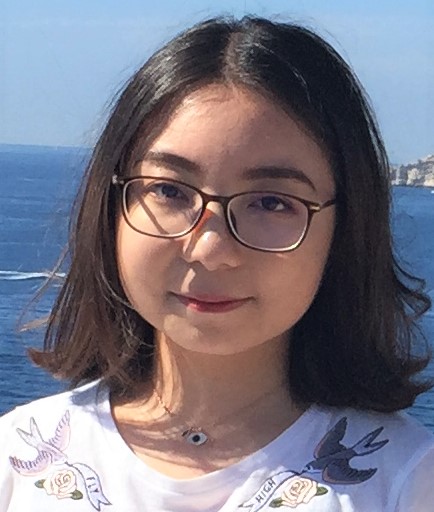}}]
{Yujin Wu}
 received the B.E. degree from the School of Electronic Engineering, Xidian University, Xi'an, China, in 2018
and the Engineering degree in computer and electronics at the Polytechnique school of Grenoble Alpes University, France in 2019. Currently, she is pursuing the Ph.D. degree in Computer science and applications at the University of Lille, France where her research interests include machine learning, multimodal emotion recognition, stress detection and computer vision.
\end{IEEEbiography}

\vspace{-14 mm}
\begin{IEEEbiography}
[{\includegraphics[width=1in,height=1.25in]{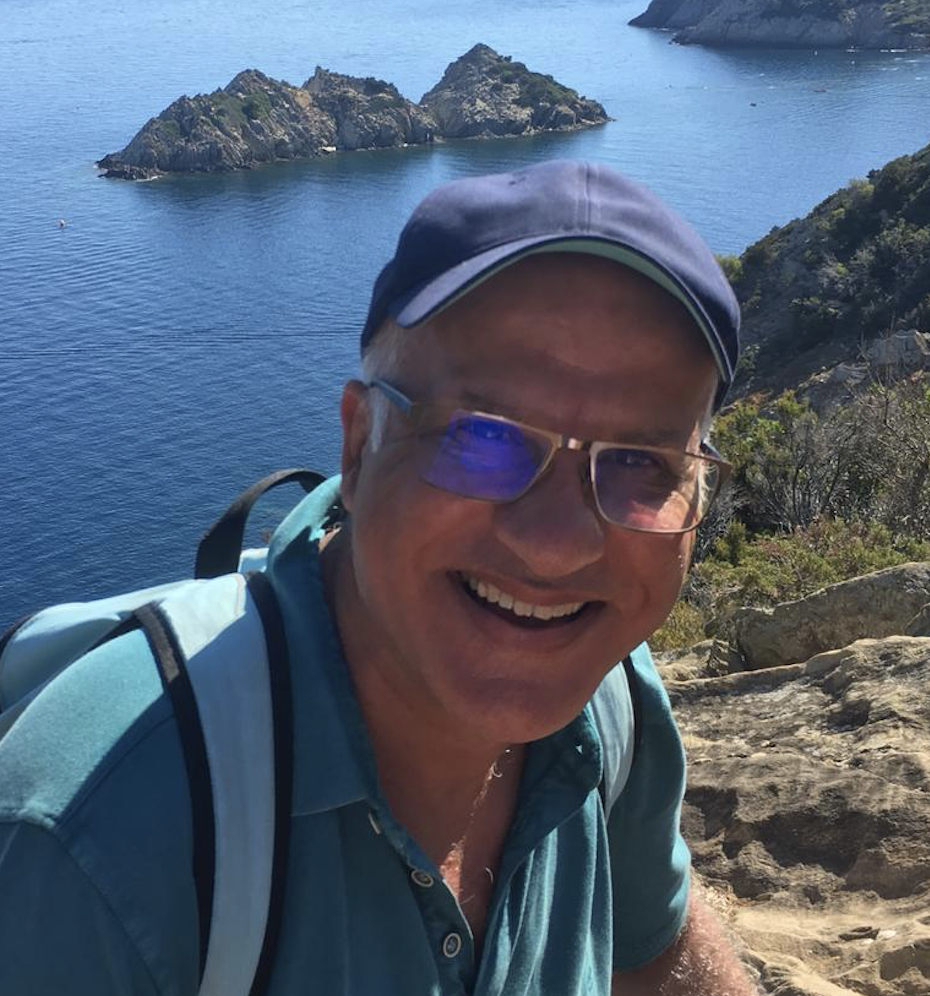}}]
{Mohamed Daoudi} is a Full Professor of Computer Science at IMT Nord Europe and the Head of Image group at CRIStAL Laboratory (UMR CNRS 9189). He received his Ph.D. degree in Computer Engineering from the University of Lille (France) in 1993. His research interests include pattern recognition, shape analysis and computer vision. He has published over 150 papers in some of the most distinguished scientific journals and international conferences. He is/was Associate Editor of Image and Vision Computing Journal, IEEE Trans. on Multimedia, IEEE Trans. On Affective Computing, and Computer Vision and Image Understanding. He has served as General Chair of IEEE International Conference on Automatic Face and Gesture Recognition, 2019. He is Fellow of IAPR and IEEE Senior member.
\end{IEEEbiography}

\vspace{-14 mm}

\begin{IEEEbiography}
[{\includegraphics[width=1in,height=1.25in]{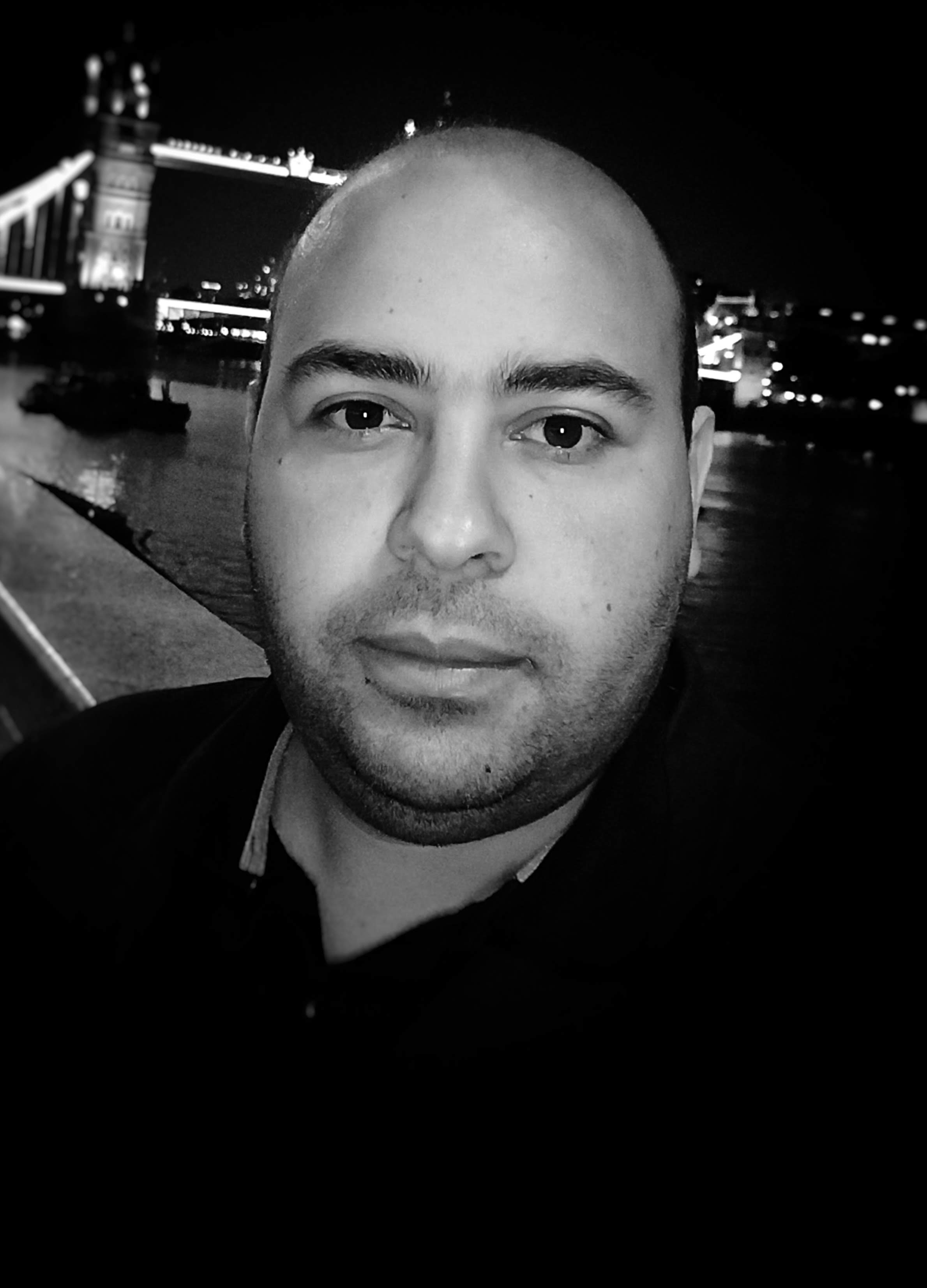}}]
{Dr. Ali Amad} is a Full Professor of psychiatry at the Lille University School of Medicine and researcher in neuroscience at Lille Neurosciences and Cognition Lab.  He received his MD degree in 2011 and his PhD degree in Neurosciences from the University of Lille in 2014. His research interests include the translation of cutting-edge research tools to everyday clinical psychiatry practice. He has published over 100 papers in scientific journals.
\end{IEEEbiography}







\end{document}